\documentclass[journal]{IEEEtran}

\usepackage{cite}
\usepackage{amsmath,amssymb,amsfonts}
\usepackage{graphicx}
\usepackage{subcaption}
\usepackage{siunitx}
\usepackage{circuitikz}
\usepackage{tikz-cd}
\usepackage{tikz}
\usepackage{enumitem}
\usepackage[firstpageonly]{draftwatermark}

\usepackage{color}
\newcommand{\REV}[1]{\textcolor{black}{{#1}}}
\newcommand{\REVV}[1]{\textcolor{black}{{#1}}}

\usetikzlibrary{shapes,arrows,positioning,calc}
\tikzstyle{block} = [draw, rectangle, minimum height=2em, minimum width=2em]
\tikzstyle{sum} = [draw, circle]
\tikzstyle{input} = [coordinate]
\tikzstyle{output} = [coordinate]
\tikzstyle{pinstyle} = [pin edge={to-,thin,black}]

\def\ddefloop#1{\ifx\ddefloop#1\else\ddef{#1}\expandafter\ddefloop\fi}

\def\ddef#1{\expandafter\def\csname c#1\endcsname{\ensuremath{\mathcal{#1}}}}
\ddefloop ABCDEFGHIJKLMNOPQRSTUVWXYZ\ddefloop

\def\ddef#1{\expandafter\def\csname s#1\endcsname{\ensuremath{\mathsf{#1}}}}
\ddefloop ABCDEFGHIJKLMNOPQRSTUVWXYZ\ddefloop

\def\ddef#1{\expandafter\def\csname b#1\endcsname{\ensuremath{\mathbf{#1}}}}
\ddefloop ABCDEFGHIJKLMNOPQRSTUVWXYZabcdefghijklmnopqrstuvwxyz\ddefloop

\def\Reals{{\mathbb R}}

\usepackage{textcomp}
\usepackage{hyperref}
\usepackage{lipsum}

\newcommand\copyrighttext{%
  \footnotesize \textcopyright 2025 IEEE. Personal use of this material is permitted.
  Permission from IEEE must be obtained for all other uses, in any current or future
  media, including reprinting/republishing this material for advertising or promotional
  purposes, creating new collective works, for resale or redistribution to servers or
  lists, or reuse of any copyrighted component of this work in other works.
  DOI: \href{https://ieeexplore.ieee.org/document/10771958}{10.1109/TCAD.2024.3509797}}
\newcommand\copyrightnotice{%
\begin{tikzpicture}[remember picture,overlay]
\node[anchor=south,yshift=10pt] at (current page.south) {\fbox{\parbox{\dimexpr\textwidth-\fboxsep-\fboxrule\relax}{\copyrighttext}}};
\end{tikzpicture}%
}

\SetWatermarkText{\textbf{SAND2024-05899O}}
\SetWatermarkAngle{0}
\SetWatermarkScale{0.07}
\SetWatermarkColor{black!}
\SetWatermarkHorCenter{0.90\paperwidth}
\SetWatermarkVerCenter{0.015\paperheight}

\begin{document}

\title{Non-intrusive data-driven model order reduction for circuits based on Hammerstein architectures
\thanks{Sandia National Laboratories is a multimission laboratory managed and operated by National Technology and Engineering Solutions of Sandia, LLC., a wholly owned subsidiary of Honeywell International, Inc., for the U.S. Department of Energy's National Nuclear Security Administration contract number DE-NA0003525.}}

\author{Joshua Hanson%
\thanks{J. Hanson is with the Department of Electrical and Computer Engineering, 
University of Illinois at Urbana-Champaign,
Urbana, IL 61801, USA (e-mail: jmh4@illinois.edu)},
\IEEEmembership{Student Member, IEEE},
\REV{Paul Kuberry}%
\thanks{\REV{P. Kuberry is with Computational Mathematics, Org. 01442,
Sandia National Laboratories, MS-1320,
Albuquerque, NM 87185-1168, USA (e-mail: pakuber@sandia.gov).}},
Biliana Paskaleva%
\thanks{B. Paskaleva is with Radiation Modeling and Analysis, Org. 01322,
Sandia National Laboratories, MS-1168,
Albuquerque, NM 87185-1168, USA (e-mail: bspaska@sandia.gov).}
\IEEEmembership{Member, IEEE}, 
Pavel Bochev%
\thanks{P. Bochev is with the Center for Computing Research,
Sandia National Laboratories, MS-1320 
Albuquerque, New Mexico, 87185-1320, USA (e-mail: pbboche@sandia.gov).}}

\maketitle
\copyrightnotice

\begin{abstract}
We demonstrate that system identification techniques can provide a basis for effective, non-intrusive model order reduction (MOR) for common circuits that are key building blocks in microelectronics.  Our approach is motivated by the practical operation of these circuits and utilizes a canonical Hammerstein architecture. To demonstrate the approach we develop  \REV{parsimonious Hammerstein models for  a nonlinear CMOS differential amplifier and an operational amplifier circuit}. We train \REV{these models} on a combination of direct current (DC) and transient Spice circuit simulation data using a novel \emph{sequential} strategy to identify \REV{their static nonlinear and linear dynamical parts}. Simulation results show that the Hammerstein model is an effective surrogate for \REV{for these types of circuits} that accurately and efficiently reproduces \REV{their} behavior over a wide range of operating points and input frequencies.
\end{abstract}


\section{Introduction}\label{sec:introduction}

Numerical circuit simulations, often referred to as Spice simulations \cite{Nagel_75_THESIS}, are foundational to the design, assessment, and qualification of modern circuits. Spice simulations use transistor-level circuit descriptions built from compact device models by application of modified nodal analysis (MNA) \cite{Ho_75_IEEETCS}. A generic compact model for a device $\sD$ is a system of differential algebraic equations (DAEs) defined by four nonlinear functions $f_{\sD,1}$, $q_{\sD,1}$, $f_{\sD,2}$, $q_{\sD,2}$:
\begin{equation}\label{eq:compact-device}
\sD \rightarrow
\left\{
\begin{aligned}
    i &= \dot{q}_{\sD,1}(t,x,v) + f_{\sD,1}(t,x,v) \\
    0 &= \dot{q}_{\sD,2}(t,x,v) + f_{\sD,2}(t,x,v)
\end{aligned}
\right. \,.
\end{equation}
In \eqref{eq:compact-device}, $v$ and $i$ are the voltages at and the currents into the device terminals, respectively, and $x$ is a vector containing internal states (e.g., charges stored in capacitors or fluxes stored in inductors). MNA applies Kirchoff’s current law at each circuit node to combine compact device models into circuit descriptions given by DAEs with the same structure as \eqref{eq:compact-device}, and size proportional to the number of devices in the circuit. As a result, Spice simulations based on transistor-level descriptions can become intractable for large circuits. For example, modern integrated circuits (IC) can have up to hundreds of millions of circuit elements resulting in large-scale DAEs with tremendous computational costs. Another complication is that the compact device models themselves, i.e., the functions $f_{\sD,1}$, $q_{\sD,1}$, $f_{\sD,2}$, $q_{\sD,2}$ need to be calibrated or modified to fit new technology and/or physics.

Computational burdens of transistor-level simulations for large circuits have spurred interest in model order reduction (MOR) for microelectronics. MOR aims to replace a highly accurate but computationally expensive \emph{full-order model} (FOM) of a system by a smaller \emph{reduced-order model} (ROM) that has acceptable accuracy and a much lower computational cost. 

One of the first and still widely used MOR approaches for circuits is \emph{macro-modeling} (MM) \cite{Ruehli_78_CAD}. A macro-model is a simplified circuit model that includes a combination of ideal circuit elements (e.g., resistors, capacitors) and dependent and independent voltage and current sources. 
The two principal MM techniques are simplification and build-up \cite{Sanchez_79_IEEETCS,Golzio_92_IJCTA}. The former successively replaces parts of the original circuit by equivalent descriptions in terms of ideal circuit elements until a sufficiently small model is obtained. Consequently, simplification yields circuit ROMs that closely resemble the original one. In contrast,  build-up constructs ad hoc blocks from ideal circuit elements that approximate the behavior of the circuit, but not its structure.
We refer to \cite{Boyle_74_IEEEJSSC,Krajewska_79_IEEEJSSC,Golzio_92_IJCTA} for further examples of MM. Although MM is applicable to general circuits, it is a manual, heuristic effort that relies on expert intuition and understanding of the circuit operation.

More rigorous and automated MOR can be developed by restricting the class of circuit FOMs to linear time-invariant (LTI) systems. These efforts have been driven primarily by the desire to improve the efficiency of post-layout simulations by reducing the size of the circuit DAE due to parasitic capacitances and resistances in interconnect structures and transmission lines. Virtually all MOR techniques for LTI systems exploit their characterization by transfer functions (TF) and achieve order reduction by replacing the ``full size'' TF by an approximate TF. 
The latter may be defined by, e.g., matching the leading terms (``moments'') of the TF's Taylor expansion at the direct current (DC) operating point, or by using the invariance of the TF with respect to equivalence transformations. The latter is the basis of balanced truncation MOR \cite[Chapter 6]{Benner_17_BOOK}, whereas the former has spawned a number of explicit and implicit moment matching approaches. Explicit techniques such as the Asymptotic Waveform Evaluation (AWE) \cite{Pillage_90_IEEETCADICC} match the moments of the TF for the FOM, which can lead to numerical issues. Implicit methods avoid these issues by first projecting the FOM onto a lower-dimensional subspace and then matching the moments of the smaller system. Typically, the projection matrix is the block Krylov basis obtained via block Arnoldi or Lanczos procedures.  The former is used in, e.g., the PRIMA  (passive reduced-order macromodels for linear RLC systems) algorithm \cite{Odabasioglu_98_IEEETCADICC}, while the second is the basis of the ROMs for linear RLC sub-circuits in \cite{Freund_00_JCAM}. We refer to \cite{Nouri_21_INPROC} for further details.

MOR  has also been developed for other specific FOM classes such as linear time-varying (LTV) systems and polynomial systems. The latter can be characterized by nonlinear ``transfer functions'' defined by Volterra kernels, which is the basis for the nonlinear model order reduction method (NORM) \cite{Li_03_INPROC,Li_05_IEEETCADICS}. NORM works by first deriving a set of minimal Krylov subspaces followed by projection of the FOM onto that set and then performing direct moment matching for the nonlinear Volterra kernels of the projected system. An example of MOR for LTV systems is the time-varying Pad\'e (TVP) method \cite{Roychowdhury_99_IEEETCSII}, which results in ROMs comprising an LTI system followed by a memoryless mixing operation, and the related approach in \cite{Ting_11_IEEEACM} for linear periodic time-varying systems.

The QLMOR (model order reduction via quadratic-linear systems) approach \cite{Gu_11_IEEETCADICS} extends moment matching to polynomializable nonlinear systems. QLMOR starts by transforming a FOM into an equivalent 
quadratic-linear DAE (QLDAE) in which dynamics are quadratic in the state variables and linear in the input variables. The QLDAE is then projected onto a lower-dimensional subspace, followed by moment-matching of the nonlinear Volterra kernels. 

Utilization of Artificial Neural Networks (ANN) as a \emph{data-driven} MOR approach for circuits is relatively new. ANNs can represent complex nonlinear behaviors by a comparatively small number of parameters, which makes them attractive for learning ROMs from data. A recent example is the ANN operational amplifier model \cite{Wei_22_IJCTA} that captures both static and dynamic behaviors of the circuit and is close to eight times faster than a transistor-level model. Another example is \cite{Xiong_21_INPROC}, which uses recurrent neural networks for transient modeling of nonlinear circuits, as well as capturing aging effects and process variations.

\REV{With the exception of ANNs}, all of these MOR techniques exploit the mathematical structure of the circuit's FOM, which has some limitations. For example, the scope of moment matching is restricted to circuits that have at most polynomial nonlinearities. QLMOR admits more general systems but at the cost of embedding them into larger, possibly higher index QLDAEs, which are more difficult to solve numerically. Reliance on FOM structure also makes these approaches \emph{intrusive} as they require access to a transistor-level description of the circuit, that may be unavailable for proprietary circuits or new technologies not yet supported by compact models. ANNs can in principle enable \emph{non-intrusive} MOR for general systems but further theoretical understanding and explainability is required for them to become a more trusted design tool.

In this paper we formulate and demonstrate an alternative system identification (SysID) MOR approach that uses the characteristic behavior of the circuit rather than its mathematical structure to guide the selection of an appropriate order reduction mechanism. \REVV{Such MOR approach can significantly accelerate early design stages and enable computationally efficient system-level modeling and analysis.}
This approach is motivated by the fact that many of the key functional building blocks in microelectronics systems such as operational amplifiers, comparators, and voltage regulators operate in a primarily memoryless capacity, and exhibit simple ``scripted'' behaviors even though their transistor-level descriptions may have rather complex nonlinear mathematical structures.
\REV{We show that such circuits can be mapped to model forms that are sufficiently expressive to capture their characteristic behaviors, yet remain simple enough to escape the computational and calibration burdens that stem from the excessive expressivity of universal model architectures.}
\REV{In so doing, we demonstrate that} SysID can provide a basis for effective and non-intrusive MOR
that is not limited by the complexity of device-level descriptions. 

\REV{Our SysID MOR follows a behavioral approach to systems modeling \cite{willems_1996}, \cite{polderman_1998}, which considers the set of admissible relationships between independent and dependent variables (e.g., input and output signals) as the central object describing a system. This perspective suggests that even very complex device-level descriptions of circuits that are characterized by ``simple'' emergent behaviors can} \REV{be compressed into} \REV{more minimal descriptions that relate independent and dependent variables using much less complex mathematical models that can serve as a basis for non-intrusive MOR for circuits.}

\REV{The focus on behaviors as the main driver for the model selection  is also one of the principal differences between this work and other applications of SysID for compact circuit modeling, such \cite{Bond_10_IEEETCDIC}. The latter formulates a universal SysID framework based on a state-space system architecture in which the state and the output equations are defined in terms of a user-selected basis. While this leads to more expressive models, it also complicates their definition and identification. Besides the basis selection, the framework in \cite{Bond_10_IEEETCDIC} requires state and basis reduction to obtain more compact models, and nonlinear constraints to enforce incremental stability. The latter leads to semidefinite programs that require sophisticated optimization algorithms. 
In contrast, by trading universality for specificity we obtain parsimonious models with simple and efficient model inference procedures that are still expressive enough to model the targeted circuit classes.} 


\REV{The rest of the paper is organized as follows. Section~\ref{sec:outline} outlines the SysID-based MOR approach and motivates the selection of the underlying model architecture. Section~\ref{sec:DiffAmp} describes the CMOS differential amplifier (DiffAmp) circuit that will be used to illustrate the key steps of the model development and inference. These steps are discussed in Section~\ref{sec:hammerstein} followed by implementation details and simulation results in Section \ref{sec:results}. Section ~\ref{sec:opamp-results} presents results for a behavioral model of a proprietary CMOS operational amplifier (OpAmp), and Section~\ref{sec:discussion} summarizes our conclusions and future work.}


\section{Outline of the approach}\label{sec:outline}

\REV{The main goal of this paper is to demonstrate that the Hammerstein model architecture, which consists of a static nonlinear map in cascade with a linear time-invariant dynamical system is an effective tool for non-intrusive MOR of circuits that operate in a predominantly memoryless capacity at low to moderate frequencies. 
Indeed, the transient dynamical behaviors of such circuits, originating from e.g., parasitic elements, exhibit small amplitudes and fast timescales that can be interpreted as  ``perturbations'' from the predominant memoryless behavior. The later can be captured by a static nonlinear block, whereas the dynamic ``perturbations'' can be approximated well in a neighborhood of the operating point by a linear system. 
In other words, the set of characteristic behaviors of such circuits can be embedded effectively into the set of behaviors expressible by a Hammerstein model.}

\REV{Furthermore, the ``separable'' nonlinear (DC) and linear (AC) components of the circuit behavior can be captured by data sets that represent the nonlinear and linear dynamics \emph{separately}. These can be identified accurately and efficiently in a \emph{sequential} manner. Such sequential approach significantly simplifies training and reduces training data and training time.}

\REV{To demonstrate  the approach we will develop Hammerstein ROMs for a small CMOS differential amplifier and a proprietary CMOS operational amplifier of a much larger size.  The first circuit will serve as a development vehicle to elucidate the key steps in the model development and the sequential model inference procedure. The second circuit will provide further evidence for the efficacy of our MOR approach. In both cases we will use similar combinations of DC and transient circuit simulation input-output data to infer the static and dynamic blocks of the models, respectively.}

Our approach falls into the category of parametric SysID methods that assume an internal model structure such as the Wiener\footnote{We note that the TVP model \cite{Roychowdhury_99_IEEETCSII} can be interpreted as a Wiener model.} \cite{Billings_82_Automatica}, Hammerstein, \cite{Kozdras_17_PE}, \cite{Gallardo_17_EC}, Hammerstein-Wiener \cite{Wills_13_Automatica}, and Wiener-Hammerstein \cite{Tan_02_IEEEIM} architectures.
\section{CMOS differential amplifier circuit}\label{sec:DiffAmp}

\begin{figure}
    \centering
    \begin{minipage}{0.48\columnwidth}
    \centering
    \resizebox{1.0\textwidth}{!}{%
    \begin{circuitikz}
        \tikzstyle{every node}=[font=\large]
        \draw [line width=0.5pt](5,9.25) to[Tnmos] (5,6.75);
        \draw [line width=0.5pt](10,11.75) to[Tnmos] (10,9.25);
        \draw [line width=0.5pt](8.75,6.75) to[Tnmos] (8.75,9.25);
        \draw [line width=0.5pt](7.5,9.25) to[Tnmos] (7.5,11.75);
        \draw [line width=0.5pt](5,14.25) to[Tpmos] (5,11.75);
        \draw [line width=0.5pt](7.5,14.25) to[Tpmos] (7.5,11.75);
        \draw [line width=0.5pt](10,11.75) to[Tpmos] (10,14.25);
        \draw [line width=0.5pt](5.85,13) to[short] (6,13);
        \draw [line width=0.5pt](5,9.25) to[short] (6.25,9.25);
        \draw [line width=0.5pt](6.25,9.25) to[short] (6.25,8);
        \draw [line width=0.5pt] (6.25,8) to[short] (6,8);
        \draw [line width=0.5pt] (6.25,13) to[short] (6,13);
        \draw [line width=0.5pt](6.25,13) to[short] (6.25,11.75);
        \draw [line width=0.5pt](5,11.75) to[short] (6.25,11.75);
        \draw [line width=0.5pt](5.85,8) to[short] (6,8);
        \draw [line width=0.5pt](7.9,8) to[short] (7.75,8);
        \draw [line width=0.5pt] (7.75,8) to[short] (6.25,8);
        \draw [line width=0.5pt](5,6.75) to[short] (8.75,6.75);
        \draw [line width=0.5pt](8.35,13) to[short] (8.5,13);
        \draw [line width=0.5pt](9.15,13) to[short] (9,13);
        \draw [line width=0.5pt](6.65,10.5) to[short] (6.5,10.5);
        \draw [line width=0.5pt](10.85,10.5) to[short] (11,10.5);
        \draw [line width=0.5pt](7.5,9.25) to[short] (10,9.25);
        \draw [line width=0.5pt](5,14.25) to[short] (10,14.25);
        \draw [line width=0.5pt](5,11.75) to[short] (5,9.25);
        \draw [line width=0.5pt](8.5,13) to[short] (9,13);
        \draw [line width=0.5pt](8.75,13) to[short] (8.75,11.75);
        \draw [line width=0.5pt] (8.75,11.75) to[short] (7.5,11.75);
        \draw [line width=0.5pt](6.75,6.75) to[short] (7,6.75);
        \draw [line width=0.5pt](7.5,14.25) to[short, -o] (7.5,15);
        \draw [line width=0.5pt](6.5,10.5) to[short, -o] (6.25,10.5);
        \draw [line width=0.5pt](11,10.5) to[short, -o] (11.25,10.5);
        \draw [line width=0.5pt](10,11.75) to[short, -o] (11.25,11.75);
        \draw [line width=0.5pt](7.5,6.75) to (7.5,6.5) node[ground]{};
        \draw (7.5,14.25) to[short, -*] (7.5,14.25);
        \draw (5,11.75) to[short, -*] (5,11.75);
        \draw (5,9.25) to[short, -*] (5,9.25);
        \draw (6.25,8) to[short, -*] (6.25,8);
        \draw (8.75,9.25) to[short, -*] (8.75,9.25);
        \draw (7.5,11.75) to[short, -*] (7.5,11.75);
        \draw (10,11.75) to[short, -*] (10,11.75);
        \draw (8.75,13) to[short, -*] (8.75,13);
        \draw (7.5,6.75) to[short, -*] (7.5,6.75);
        \node [font=\large] at (4.75,8) {$M_6$};
        \node [font=\large] at (9,8) {$M_5$};
        \node [font=\large] at (7.75,10.5) {$M_1$};
        \node [font=\large] at (9.75,10.5) {$M_2$};
        \node [font=\large] at (10.25,13) {$M_4$};
        \node [font=\large] at (4.75,13) {$M_7$};
        \node [font=\large] at (7.25,13) {$M_3$};
        \node [font=\large] at (6.25,11) {$V_1$};
        \node [font=\large] at (11.25,11) {$V_2$};
        \node [font=\large] at (11.25,12.25) {$V_3$};
        \node [font=\large] at (7.5,15.5) {$V_{\text{DD}}$};
    \end{circuitikz}}
    \caption{\small  Schematic of the CMOS nonlinear differential amplifier circuit considered in this work \cite{Prasad_16_report}. The body of each NMOS transistor is connected to ground, and the body of each PMOS transistor is connected to $V_{\text{DD}}$.}
    \label{fig:CMOS-diff-amp}
    \end{minipage}
    \hfill
    \begin{minipage}{0.48\columnwidth}
    \centering
    \resizebox{1.0\textwidth}{!}{%
    \begin{circuitikz} 
        \tikzstyle{every node}=[font=\large]
        \ctikzset{resistor = european}
        \draw (6.75,14) to[R,l={ \large $Z_{\text{in}}$}] (6.75,11);
        \draw (8.25,13) to[R,l={ \large $Z_{\text{out}}$}] (10.75,13);
        \draw (8.25,12.75) to[american controlled voltage source,l={ \large $A_{\text{d}}(V_1-V_2)$}] (8.25,11.25);
        \draw (8.25,11.5) to (8.25,11.25) node[ground]{};
        \draw [](6.75,14) to[short, -o] (5.25,14);
        \draw [](6.75,11) to[short, -o] (5.25,11);
        \draw [short] (12.5,12.5) -- (6.25,15.75);
        \draw [](10.75,12.5) to[short, -o] (13.25,12.5);
        \draw [short] (6.25,15.75) -- (6.25,9.25);
        \draw [short] (6.25,9.25) -- (12.5,12.5);
        \draw [](10.75,13) to[short] (10.75,12.5);
        \node [font=\large] at (5.25,14.5) {$V_1$};
        \node [font=\large] at (5.25,11.5) {$V_2$};
        \node [font=\large] at (13.25,13) {$V_3$};
        \draw [](8.75,14.45) to[short, -o] (8.75,15); 
        \draw [](8.75,10) to[short, o-] (8.75,10.55); 
        \node [font=\large] at (8.75,15.5) {$+V$}; 
        \node [font=\large] at (8.75,9.5) {$-V$}; 
        \draw [](8.25,13) to[short] (8.25,12.75);
    \end{circuitikz}}
    \caption{\small A basic linear differential amplifier model. The impedances $Z_{\text{in}}$ and $Z_{\text{out}}$ may in general comprise both resistive and reactive parts. The supply voltages $+V$ and $-V$ establish upper and lower bounds on the voltage produced by the dependent source.}
    \label{fig:basic-diff-amp}
    \end{minipage}
\end{figure}

The main function of an analog differential amplifier is to produce an output voltage $V_3$ that is proportional to the difference between two input voltages $V_1$ and $V_2$. A basic memoryless linear differential amplifier can thus be modeled by the equation $V_3 = A_{\text{d}}(V_1-V_2)$, where $A_{\text{d}}$ is a large positive constant called the \textit{differential gain}; see Figure~\ref{fig:basic-diff-amp}. An additional non-ideal term $A_{\text{cm}} (V_1 + V_2)$ could be included on the right-hand side to capture the unintended amplification of the average value of the input voltages, where the constant $A_{\text{cm}}$ is called the \textit{common-mode gain}. In general we can express the output voltage as a global nonlinear function of both of the input voltages.

When the ports of this circuit are connected to non-ideal sources or load, current flowing through the input and output impedances will create a discrepancy between the ideal input and output voltages and those actually produced at the terminals. This discrepancy is determined by the relationship between the input and output impedances of the external elements and the internal input and output impedances of the circuit. We typically represent these non-ideal input and output impedances using some combination of resistors and capacitors, yielding a linear time-invariant sub-system. Replacing the expression controlling the dependent source with a nonlinear function of the input voltages, the basic differential amplifier model in Figure~\ref{fig:basic-diff-amp} can be represented exactly in the form of a Hammerstein model, with the port voltages assigned to the input variables of the model and the port currents assigned to the output variables.

The nonlinear relationship between the input and output voltages originates from the voltage-dependent drain (resp. source) currents in the NMOS (resp. PMOS) transistors. In the Shichman-Hodges MOSFET model \cite{Shichman_68_IEEEJSSC}
this is represented by a piecewise-quadratic nonlinear function. The parasitic capacitances between transistor terminals are responsible for current flow during transient operation. The interplay between these constituent elements yields the emergent behavior of the overall differential amplifier.

When the circuit in Figure \ref{fig:CMOS-diff-amp} is included as a sub-circuit in the context of a larger system, a global Spice simulation only requires information about the sub-circuit behavior at the ports. Thus the aim of a reduced-order model is to subsume all of the individual transistor current-voltage characteristics into a unified expression relating only the voltages and currents at the external ports, which allows us to abstract away the detailed internal behaviors. In this way, intermediate computations for internal node voltages and branch currents are eliminated, yielding a more computationally efficient model.

The differential amplifier considered in this paper follows the analysis presented \cite{Prasad_16_report}, and was designed to meet a specific set of operational characteristics. In particular, the widths and the lengths of the individual transistors were selected in order for the circuit to meet the specifications in 
\cite[Tables 1-2]{Prasad_16_report}. Several of the circuit's dynamical characteristics such as the cutoff frequency $f_{-3 \text{dB}}$ and slew rate are derived for a specific load connected to the output terminal. For this specific design analysis the load capacitor  $C_{\text{L}}$ is set to 5 pF.


\section{Hammerstein Model}\label{sec:hammerstein}

Model structure selection is one of the cornerstones of system identification. If the architecture chosen is incapable of exhibiting the behaviors of interest, then the model accuracy will suffer regardless of the training stimulus and optimization algorithms used. On the other hand if the architecture is extremely expressive, the model will likely be capable of accurately reproducing the training data, but it may incur excessive computational burden to evaluate and possibly demonstrate poor generalization to new data. In the context of non-intrusive reduced-order modeling, we seek the simplest possible model structure that remains expressive enough to accurately manifest the desired behaviors.


\REV{As explained in Section \ref{sec:outline}, Hammerstein models --- represented by a static nonlinearity in cascade with a linear time-invariant system; see Figure \ref{fig:generic-hamm-model} --- are good candidates for emulating the behavior of circuits that operate in a predominately memoryless capacity.}
In a certain sense, this is the simplest possible nonlinear model architecture suited for capturing mainly ``DC'' behavior and correcting for parasitic and loading effects with an ``AC'' augmentation. Th\'evenin and Norton equivalent circuits for linear electrical networks share a similar structure. To form a Hammerstein model from a Th\'evenin or Norton equivalent circuit, the lumped voltage or current source is replaced by a dependent source with a nonlinear relationship to some independent input variables (usually port voltages), and the lumped impedance serves the role of the dynamic linear block. 

\begin{figure}
    \centering
    \resizebox{0.45\textwidth}{!}{%
    \begin{circuitikz}
        \tikzstyle{every node}=[font=\Huge]
        \draw [line width=1.5pt] (7.5,14.25) rectangle node {\Huge $\varphi$} (10,11.75);
        \draw [line width=1.5pt] (12.5,14.25) rectangle node {\Huge $H$} (15,11.75);
        \node [font=\Huge] at (5,13.75) {$u$};
        \node [font=\Huge] at (17.5,13.75) {$y$};
        \draw [line width=1.5pt, ->] (5,13) -- (7.5,13);
        \draw [line width=1.5pt, ->] (10,13) -- (12.5,13);
        \draw [line width=1.5pt, ->] (15,13) -- (17.5,13);
    \end{circuitikz}}
    \caption{\small  Block diagram of the generic Hammerstein model structure, where $\varphi$ is a memoryless nonlinear function, $H$ is a linear time-invariant system, $u$ is the input, and $y$ is the output.}
    \label{fig:generic-hamm-model}
\end{figure}

\subsection{Multi-port circuit models}

For input $u$ and output $y$ (which may in general be vector-valued), the canonical state-space representation of the generic Hammerstein model in Figure~\ref{fig:generic-hamm-model} takes the form
\begin{equation}\label{eq:hammerstein}
\begin{split}
    \dot{x} &= Ax + B\varphi(u) \\
    y &= Cx + D\varphi(u),
\end{split}
\end{equation}
where $A,B,C,D$ are matrices and $\varphi$ is a nonlinear map. Although generic behavioral circuit models do not require explicitly distinguishing between input and output variables, for convenience of integration into standard circuit simulation software we typically identify port voltages as the input variables and port currents as the output variables. Such systems can be written as a system of differential-algebraic equations
\begin{equation*}
\begin{split}
    i &= \dot{q}_1(t,x,v) + f_1(t,x,v) \\
    0 &= \dot{q}_2(t,x,v) + f_2(t,x,v),
\end{split}
\end{equation*}
which has the same structure as the generic compact device model in \eqref{eq:compact-device}. As in that model, $v$ denotes the port voltages, $i$ denotes the port currents, and internal states (e.g., charges stored in capacitors or fluxes stored in inductors) are subsumed in $x$. To implement the Hammerstein model in this standard form, set $u \gets v$, $y \gets i$ and define
\begin{alignat*}{3}
    q_1(t,x,v) &= 0, \quad & f_1(t,x,v) &= Cx + D\varphi(v) \\
    q_2(t,x,v) &= -x, \quad & f_2(t,x,v) &= Ax + B\varphi(v).
\end{alignat*}
For the purposes of training the model, we can work directly with \eqref{eq:hammerstein} and convert the identified model into the proper form afterwards, avoiding the need to solve any implicit DAEs within the training loop.

We now proceed to specialize the generic Hammerstein model in Figure~\ref{fig:generic-hamm-model} to develop a ROM for the CMOS differential amplifier circuit. Since the linear block in this model is a standard LTI system, the  key step in the specialization process is the design of the static nonlinear block $\varphi$. 

\subsection{Design of the static nonlinearity}\label{ssec:nonlinearity_design}

To define the nonlinear map $\varphi$ we shall take advantage of the fact that the circuit in Figure \ref{fig:CMOS-diff-amp} is approximately memoryless, which also holds for many other similar small amplifier topologies. This behavior suggests that the transient port currents ${(i_1,i_2,i_3) : [0,T] \to \Reals^3}$ resulting from the time-varying port voltages ${(v_1,v_2,v_3) : [0,T] \to \Reals^3}$, evaluated at a given time $t \in [0,T]$, can be roughly approximated by the DC port currents $(I_1,I_2,I_3)$ resulting from the fixed DC port voltages $(V_1,V_2,V_3) = (v_1(t),v_2(t),v_3(t))$. In this circuit, the difference between $(I_1,I_2,I_3)$ and $(i_1(t),i_2(t),i_3(t))$ can be compensated by an attenuation and phase shift characteristic of a linear filter in addition to a subtle frequency-dependent bias shift and harmonic distortion.

\begin{figure}
    \centering
    \resizebox{0.45\textwidth}{!}{%
    \begin{circuitikz}
        \tikzstyle{every node}=[font=\Huge]

        \draw[dashed] (-0.5,2.5)--(-0.5,14.75);
        \draw[dashed] (-0.5,14.75)--(10.25,14.75);
        \draw[dashed] (10.25,14.75)--(10.25,2.5);
        \draw[dashed] (10.25,2.5)--(-0.5,2.5);

        \draw[dashed] (10.75,2.5)--(10.75,14.75);
        \draw[dashed] (10.75,14.75)--(26.75,14.75);
        \draw[dashed] (26.75,14.75)--(26.75,2.5);
        \draw[dashed] (26.75,2.5)--(10.75,2.5);

        \draw [line width=2pt] (0,14.25) rectangle  node {\Huge $I_{\text{DC}}$} (2.5,11.75);
        \draw [line width=2pt] (5,11.75) rectangle  node {\Huge $I_{\text{DC}}^2$} (7.5,9.25);
        \draw [line width=2pt] (11.25,14.25) rectangle  node {\Huge $B$} (13.75,9.25);
        \draw [line width=2pt] (11.25,8) rectangle  node {\Huge $D$} (13.75,3);
        \draw [line width=2pt] (16.25,11.75) circle (1.25cm) node {\Huge $+$} ;
        \draw [line width=2pt] (18.75,13) rectangle  node {\Huge $\int$} (21.25,10.5);
        \draw [line width=2pt] (18.75,9.25) rectangle  node {\Huge $A$} (21.25,6.75);
        \draw [line width=2pt] (25,5.5) circle (1.25cm) node {\Huge $+$} ;
        \draw [line width=2pt] (23.75,13) rectangle  node {\Huge $C$} (26.25,10.5);
        \node [font=\Huge] at (-2.5,14.00) {$(v_1,v_2,v_3)$};
        \node [font=\Huge] at (28.75,6.5) {$(i_1,i_2,i_3)$};
        \draw [line width=2pt, ->] (2.5,13) -- (11.25,13);
        \draw [line width=2pt, ->] (21.25,11.75) -- (23.75,11.75);
        \draw [line width=2pt, ->] (25,10.5) -- (25,6.75);
        \draw [line width=2pt, ->] (16.25,8) -- (16.25,10.5);
        \draw [line width=2pt, ->] (22.5,8) -- (21.25,8);
        \draw [line width=2pt, ->] (7.5,10.5) -- (11.25,10.5);
        \draw [line width=2pt, ->] (9.75,6.75) -- (11.25,6.75);
        \draw [line width=2pt, ->] (8.5,4.25) -- (11.25,4.25);
        \draw [line width=2pt, ->] (3.75,10.5) -- (5,10.5);
        \draw [line width=2pt, ->] (13.75,11.75) -- (15,11.75);
        \draw [line width=2pt, ->] (17.5,11.75) -- (18.75,11.75);
        \draw [line width=2pt, ->] (13.75,5.5) -- (23.75,5.5);
        \draw [line width=2pt, ->] (26.25,5.5) -- (28.75,5.5);
        \draw [line width=2pt, ->] (-2.5,13) -- (0,13);
        \draw [line width=2pt, short] (3.75,13) -- (3.75,10.5);
        \draw [line width=2pt, short] (8.5,10.5) -- (8.5,4.25);
        \draw [line width=2pt, short] (9.75,13) -- (9.75,6.75);
        \draw [line width=2pt, short] (22.5,11.75) -- (22.5,8);
        \draw [line width=2pt, short] (18.75,8) -- (16.25,8);
    \end{circuitikz}}
    \caption{\small  Block diagram of the proposed Hammerstein model for the CMOS DiffAmp circuit. \REVV{The left dashed-outline section depicts the nonlinearity $\varphi$ as described in Section \ref{ssec:nonlinearity_design}; the right dashed-outline section depicts a general linear time-invariant system.}}
    \label{fig:CMOS-hamm-model}
\end{figure}

We can exploit this knowledge to design an effective nonlinearity $\varphi$ as follows. Let
$ I_{\text{DC}} : \Reals^3 \to \Reals^3$,  $(V_1,V_2,V_3) \mapsto (I_1,I_2,I_3)$
denote the ${I\!-\!V}$ characteristics of the circuit, i.e., the map that computes the DC port currents resulting from a given set of DC port voltages. We then define the map
\begin{equation}\label{eq:NL}
    \varphi : \Reals^3 \to \Reals^3 \times \Reals^3,\quad (V_1,V_2,V_3) \mapsto (I_{\text{DC}},I_{\text{DC}}^2),
\end{equation}
as a composition of component-wise identity and squaring operations with the ${I\!-\!V}$ map $I_{\text{DC}}$ of the circuit. The squaring operation provides a prototypical nonlinearity that, when appropriately filtered by a linear system, enables the model to capture the subtle bias shift and harmonic distortion introduced at higher frequencies. Implementation of the nonlinear block \eqref{eq:NL} requires a suitable approximation of the nonlinear map $I_{\text{DC}}$. There are several options that one can consider for this purpose such as parameterizing $I_{\text{DC}}$ as a member of some generic expressive function class (e.g., polynomials, rational functions, neural nets), \REV{some of which are especially well-suited for handling high-dimensional data,} or using non-parametric regression such as Moving Least Squares (MLS); see, e.g., \cite{Bochev_20_PMLR} for applications of neural net and MLS regression of ${I\!-\!V}$ characteristics to develop compact device models. Here we choose to represent $I_{\text{DC}}$ by using a table-based approach \cite{Gupta_18_THESIS,Gupta_17_INPROC} implemented with piecewise-trilinear \REV{(Q3)}
interpolants. To that end, assume that the full range of possible port voltages is given by the box\footnote{\REV{For handling higher-dimensional data produced by e.g., circuits with a large number of ports, one may elect to implement some method of sparse or adaptive sampling, rather than dense hypercube sampling, to reduce data requirements while still enabling interpolation-based function classes.}}
\begin{equation}\label{eq:box}
    D = \prod_{i=1}^{3} [V_i^{\min},V_i^{\max}] \subset \Reals^3\,,
\end{equation}
where $V_i^{\min}$ and $V_i^{\max}$ denote lower, resp. upper bounds on the voltage $V_i$ at the $i$th port. Let $K_i>1$ be an integer specifying the sampling density for the $i$th port. We then perform a DC sweep using $K_i$ voltage points ${V}_i^{\min} \le \widehat{V}_{i,k_i}\le V_i^{\max}$, $k_i=1,\ldots,K_i$ for each of the ports, to obtain a total of $K=K_1\times K_2\times K_3$  voltage-current pairs $\{(\widehat{V}_{1,k_1},\widehat{V}_{2,k_2},\widehat{V}_{3,k_3}),(\widehat{I}_{1,k_1},\widehat{I}_{2,k_2},\widehat{I}_{3,k_3})\}$. The sample voltages $(\widehat{V}_{1,k_1},\widehat{V}_{2,k_2},\widehat{V}_{3,k_3})$ form a Cartesian grid on $D$, however, this grid is not required to be uniform.

We then approximate the map $I_{\text{DC}}$ by the  \REV{Q3} interpolant $I^h_{\text{DC}}$ of the voltage-current pairs defined above. Computation of $I_{\text{DC}}$ for a given set of DC port voltages $V=(V_1,V_2,V_3)$ is a local operation that only requires data from the Cartesian grid cell $\widehat{D}(V)$ containing $V$. To explain this step, denote for notational clarity the vertices of this cell as $\widehat{V}_{pqr}=(\widehat{V}_{1,p},\widehat{V}_{2,q},\widehat{V}_{3,r})$ with $p,q,r=0,1$ and let $\psi_{i,\sigma}$, $i=1,2,3$, $\sigma=0,1$ denote the local linear basis functions such that $\psi_{i,\sigma}(\widehat{V}_{i,s}) = \delta_{\sigma,s}$ for $\sigma,s=0,1$. Then the port current at the $j$th port corresponding to the input port voltage $V$ can be approximated by
\begin{equation}\label{eq:Q1}
    I^h_{j} = \sum_{p,q,r=0,1} \widehat{I}_{j,pqr} \psi_{1,p}(V_1)\psi_{2,q}(V_2)\psi_{3,r}(V_3)\,
\end{equation}
where $\widehat{I}_{j,pqr}$ is the current at the $j$th port corresponding to vertex $\widehat{V}_{pqr}$. One can show \cite{Ciarlet_02_BOOK} that the map $I_{\text{DC}}$ defined by \eqref{eq:Q1} is second-order accurate in the sense that
\begin{equation}\label{eq:error}
    | I^h_{j} - I_{j}(V) | \le C \text{diam}(\widehat{D}(V))^2 \sup_{\hat{V}\in \widehat{D}}| D^2 I_j(\hat{V})|\,,
\end{equation}
where $I_{j}(V)$ is the exact port current corresponding to $V$, $C$ is a positive constant, and $D^2$ is the total second derivative operator. This error bound implies that approximation error will be larger in regions where $I_j$ has steep gradients. We note that when $V$ coincides with a grid node $\widehat{V}_{pqr}$ formula \eqref{eq:Q1} simply returns the corresponding port currents $ \widehat{I}_{j,pqr}$. The memory requirements and the accuracy of this approach are directly proportional to the number of voltage-current pairs $K=K_1\times K_2\times K_3$. Increasing $K$ improves the error bound \eqref{eq:error} but could lead to a sizeable memory footprint. Nonetheless, trilinear interpolation is simpler to set up, cheaper to evaluate, and requires less storage than a cubic or higher order spline-based interpolation scheme, which is the primary reason we use \eqref{eq:Q1} in this paper.

An important consequence of a table-based regression of the ${I\!-\!V}$ characteristic is that it effectively leads to a map $\varphi$ that has no unidentified parameters --- all model degrees-of-freedom are pushed entirely into the linear filter, which is responsible for shaping the raw output $(I_{\text{DC}},I_{\text{DC}}^2)$ into the transient currents $(i_1,i_2,i_3)$. This fact forms the basis of our sequential parameter identification strategy, which we present in the next section.

\subsection{Sequential parameter identification algorithm}

Consider again the circuit in Figure \ref{fig:CMOS-diff-amp}. Suppose we connect terminals $1,2,3$ to some external voltage sources and loads, which should be chosen to reproduce typical operating conditions for this circuit. The behaviors we want the model the discover should be induced by the training stimulus. Simulating or measuring this circuit yields time-series data for the port voltages ${(v_1(t_j),v_2(t_j),v_3(t_j))_{j=1}^N}$ and port currents ${(i_1(t_j),i_2(t_j),i_3(t_j))_{j=1}^N}$ where ${\{t_j\}_{j=1}^N \subset [0,T]}$. One may perform several experiments and aggregate the data, or perform a single comprehensive experiment and collect a single time-series; in this work we adopt the latter approach.

The standard parameter identification problem for \eqref{eq:hammerstein} then is to produce a map $\varphi$ and matrices $A,B,C,D$ such that for a given input $u = (v_1,v_1,v_3)$, the error between the model output $y$ and the training output $(i_1,i_2,i_3)$ is minimized. A common choice, also used in this work, is to measure this error in the $L^2$ norm $\|\cdot\|$ on the time-interval $[0,T]$, i.e., we consider the following Mean-Squared Error (MSE) loss functional
\begin{equation}\label{eq:loss}
J(y,i) = \frac{1}{2} \| y - (i_1,i_2,i_3) \|^2 \,.
\end{equation}
However, as explained in the previous subsection, we construct the map $\varphi$ directly from DC sweep data using table-based piecewise-trilinear interpolation and so, $\varphi$ does not include any free parameters. As a result, the only degrees-of-freedom that are left to be identified in our model are the matrices $A,B,C,D$  in \eqref{eq:hammerstein}. Let $n$ be a positive integer specifying the internal state dimension of the dynamic block. The number of inputs to this linear block is equal to six (three for $I_{\text{DC}}$ and three for $I_{\text{DC}}^2$ provided by $\varphi$), which yield matrix dimensions $A \in \Reals^{n \times n}$, $B \in \Reals^{n \times 6}$, $C \in \Reals^{3 \times n}$, $D \in \Reals^{3 \times 6}$.

To find these matrices we solve the following constrained optimization problem: 
\begin{equation}\label{eq:opt}
    \mbox{\emph{Find} $A,B,C,D$ \emph{to minimize \eqref{eq:loss} subject to \eqref{eq:hammerstein}.}}
\end{equation}
We call this model inference procedure \emph{sequential parameter identification} because in contrast to conventional identification of Hammerstein models, the nonlinear block $\varphi$ is defined directly and \emph{independently} from the linear block by using DC sweep data. As a result, the inference of \eqref{eq:hammerstein} is effectively reduced to identification of an LTI system, which significantly simplifies the training of the model and reduces the amount of training data required. We reiterate that the appropriateness of this sequential parameter identification procedure stems from the fact that our model circuit is approximately memoryless.

Let us now discuss the second stage of this procedure, i.e., the identification of the matrices $A,B,C,D$ in \eqref{eq:hammerstein} with particular emphasis on the generation of time-series training data for the optimization problem \eqref{eq:opt}. We recall that this data should correspond to common operating conditions for our circuit, such as those described in Section \ref{sec:DiffAmp}. To that end we consider a typical configuration of the CMOS amplifier obtained by connecting a 5 pF capacitive load to the output port at terminal 3, and two voltage sources to the gates at terminals 1 and 2. We then simulate the circuit in Xyce \cite{Keiter_20_TECHREPORT} using voltage sources producing time-varying waveforms given by the following frequency-modulated sinusoids (also known as exponential \textit{chirp} signals \cite{Novak_10_IEEE}):
\begin{equation}\label{eq:training_voltages}
\begin{split}
    v_1(t) &= V_{\text{bias}} + A \sin \left( \phi(t) \right) \\
    v_2(t) &= V_{\text{bias}} - A \sin \left( \phi(t) \right),
\end{split}
\end{equation}
where the phase is given by
\begin{equation*}
    \phi(t) = 2 \pi \int_0^T f_0^{1 - \frac{\tau}{T}} f_1^{\frac{\tau}{T}} \mathop{}\!\mathrm{d} \tau = 2 \pi \frac{ f_0^{1 - \frac{t}{T}} f_1^{\frac{t}{T}} - f_0 }{\ln(f_1) - \ln(f_0)} T.
\end{equation*}
The instantaneous frequency $f(t) = f_0^{1 - \frac{t}{T}} f_1^{\frac{t}{T}}$ sweeps from $f_0$ at time $0$ to $f_1$ at time $T$. The parameters are chosen according to the following:
\begin{itemize}
    \item The high frequency $f_1 = 5$ GHz is chosen based on the intended operating regime of the circuit.
    \item The low frequency $f_0 = 100$ kHz is chosen sufficiently small to be dominated by characteristic time scale of the circuit; on the other hand, choosing a value closer to $f_1$ will necessitate fewer periods, reducing simulation time.
    \item The bias $V_{\text{bias}} = 2.5$ V is the nominal DC operating point given by the midpoint of $V_{\text{DD}}$ and ground.
    \item The amplitude $A = 50$ mV is large enough to draw out the frequency dependent bias shift and harmonic distortion without introducing exaggerated hard clipping.
    \item The number of periods is chosen to be $N_{\text{per}} = 100$. The time horizon is given by $T = N_{\text{per}} \frac{\ln(f_1) - \ln(f_0)}{f_1 - f_0} \approx 216$ ns.
\end{itemize}
We sample these waveforms using a variable time step in such a way that each period is sampled using an equal number of points, in this case 500 points per period. The transient simulation in Xyce produces time-series for the time-varying port currents $\bi_N=(i_1(t_j),i_2(t_j),i_3(t_j))_{j=1}^N$ and port voltages $\bv_N=(v_1(t_j),v_2(t_j),v_3(t_j))_{j=1}^N$, where $v_1$ and $v_2$ are specified in \eqref{eq:training_voltages} and $v_3$ is computed from the transient simulation of the circuit with the capacitive load. Thus the input to the linear system \eqref{eq:hammerstein}  we wish to identify consists of the output from the static nonlinearity $\varphi$ evaluated on the time-varying port voltages $\bv_N$, i.e.,
\begin{align*}
    \bI_N &= (I_1(t_j),I_2(t_j),I_3(t_j), I_1^2(t_j),I_2^2(t_j),I_3^2(t_j)) \\
    & := \varphi(v_1(t_j),v_2(t_j),v_3(t_j))
\end{align*}
for $j=1,\ldots,N$, while the target output is the time-varying port currents $\bi_N$. Numerical solution of the optimization problem \eqref{eq:opt} then yields matrices $A,B,C,D$ that minimize the MSE between the output $y$ of the linear system \eqref{eq:hammerstein} and the transient port currents $\bi_N$ computed using Xyce.

The choice of training stimulus is important because the range of validity of the model is limited by what is represented in the simulation data from the original circuit. In this case, this comprises input voltage waveforms that are between 0 and 5 V and bandlimited to approximately 4 GHz, with capacitive loads connected to the output port. The exponentially-modulated frequency in the waveforms above is designed so that a more evenly distributed amount of time is spent in each frequency regime. At very low frequencies, the model behavior by construction reproduces the nonlinear map $\varphi$, whose accuracy can be controlled by the user through a selection of the DC sampling density, i.e., the integers $K_i$ specifying the numbers of sampling points for each input port.

To identify an $n$-dimensional linear state space model, it suffices for the input to have $n$ distinct frequencies. The waveform described above contains in a certain sense an uncountable number of frequencies, resembling an overdetermined regression problem. To explain this choice consider the following analogy. Suppose we have a dataset $X$ comprising points $\{(x_i,y_i)\}_{i=1}^{P}$, which we know satisfy an approximately linear relationship.  Since a line in the Euclidean plane is uniquely determined by only two points we could build a linear model for $X$ based on just two samples from this dataset. This approach would provide a perfect fit for two of the samples, but the rest of $X$ may not be approximated equally well. Identifying the line using linear regression yields a better overall fit to $X$ by instead minimizing the sum of squared residuals to more than two points.

In the dynamic setting, we conjecture that the underlying system that relates the outputs $\bI_N$ of $\varphi$ to the transient port currents $\bi_N$ is approximately linear. We can construct an overdetermined regression problem to achieve a more robust fit by including more frequencies in the training data than is strictly necessary to uniquely identify a linear system of the desired state dimension, which is precisely the aim of the frequency-modulated input waveforms in \eqref{eq:training_voltages}.

\section{Simulation results \REV{with differential amplifier}}\label{sec:results}
In this section we compare simulation results for the CMOS differential amplifier \REV{(DiffAmp)} in Figure \ref{fig:CMOS-diff-amp} and a Hammerstein model constructed and identified according to Section \ref{sec:hammerstein}. The simulation results for the circuit, which we consider as ground truth, are computed using Sandia's analog circuit simulator Xyce \cite{Keiter_20_TECHREPORT}. 

\subsection{Implementation and setup}\label{sec:impl}
We implemented the reduced-order Hammerstein model in Python, with the transient simulations utilizing the Jax-based differential equations library Diffrax \cite{Kidger_21_PHDTHESIS}. To identify the parameters $A,B,C,D$ of the Hammerstein model \eqref{eq:hammerstein} we solve \eqref{eq:opt} using a quasi-Newton reduced space approach \cite{Hicken_13_AIAA}. We employ the limited-memory Broyden–Fletcher–Goldfarb–Shanno (BFGS) \cite{Fletcher_00_BOOK} algorithm and use he automatic differentiation capability from Diffrax~\cite{Kidger_21_PHDTHESIS} to compute the gradients of \eqref{eq:loss}.

When presenting results we will discard the gate currents $i_1$ and $i_2$ because they are exactly zero (in the DC results) or nearly identically zero (in the transient results). Instead we only showcase the output voltage $v_3$ and output current $i_3$ resulting from (constant or time-varying) input voltages $v_1$ and $v_2$. The input terminals 1 and 2 will always be connected to independent voltage sources and the output terminal 3 will either be connected to another independent source (in the DC results) or a 5 pF capacitive load (in the transient results).

\subsection{DC results}\label{sec:DC_results}
In this subsection we assess the accuracy of the nonlinear input block $\varphi$ of the Hammerstein model \eqref{eq:hammerstein}. For circuits that are approximately memoryless, this block provides a simplified version of \eqref{eq:hammerstein} that is appropriate for relatively low-frequency signals. In such cases one can choose to run the Hammerstein model without the dynamic block.

The following results are derived from an iterated DC operating point analysis, which solves for the DC port currents $(I_1,I_2,I_3)$ when a given tuple of port voltages $(V_1,V_2,V_3)$ are applied. To that end, we define the DC sweep box $D$ in \eqref{eq:box} by setting $V_i^{\min}=0$ and $V_i^{\max}=5$ for $i=1,2,3$ and sweep each voltage with a step size of $h = 100$ mV. This step corresponds to a sampling density $K_i = 5001$ for all input voltages. 

We use this Xyce data to construct the nonlinear map $\varphi$ by piecewise-trilinear interpolation as described in Section \ref{sec:hammerstein}. We then report the DC output current $I^h_3$ of the nonlinear block as a function  of the two input voltages $V_1$ and $V_2$ applied to the gate terminals 1 and 2, respectively, while $V_3$ is being held fixed. The accuracy of this current can be estimated using the error bound \eqref{eq:error}. Since we sample uniformly along each direction of the DC sweep box \eqref{eq:box}, the diameter of each grid cell is $d = \sqrt{3} h$, where $h=0.001V$. Thus,
\begin{equation}\label{eq:err-est-2}
    |I^h_3 - I_3({V}) | \le \sup_{\hat{V}\in \widehat{D}}| D^2 I_3(\hat{V})|\times 10^{-6} \,.
\end{equation}
where $\widehat{D}$ is the cell containing $V$. Similar bound holds for $V_3$.

\begin{figure}
    \centering
    \begin{subfigure}{0.5\textwidth}
      \centering
      \includegraphics[trim={2cm 0.6cm 2cm 2cm},clip,width=0.95\linewidth]{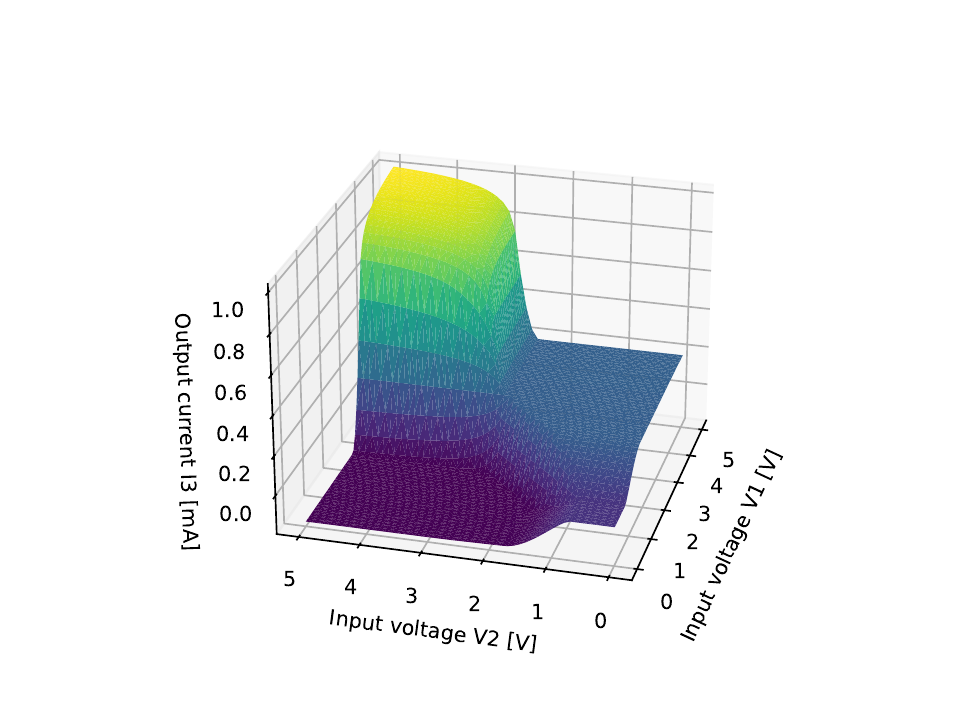}
      \caption{\small  DC output current $I^h_3(V_1,V_2)$ for $V_3$ = 1.5 V.}
      \label{fig:sub1}
    \end{subfigure}
    \begin{subfigure}{0.5\textwidth}
      \centering
      \includegraphics[trim={2cm 0.6cm 2cm 2cm},clip,width=0.95\linewidth]{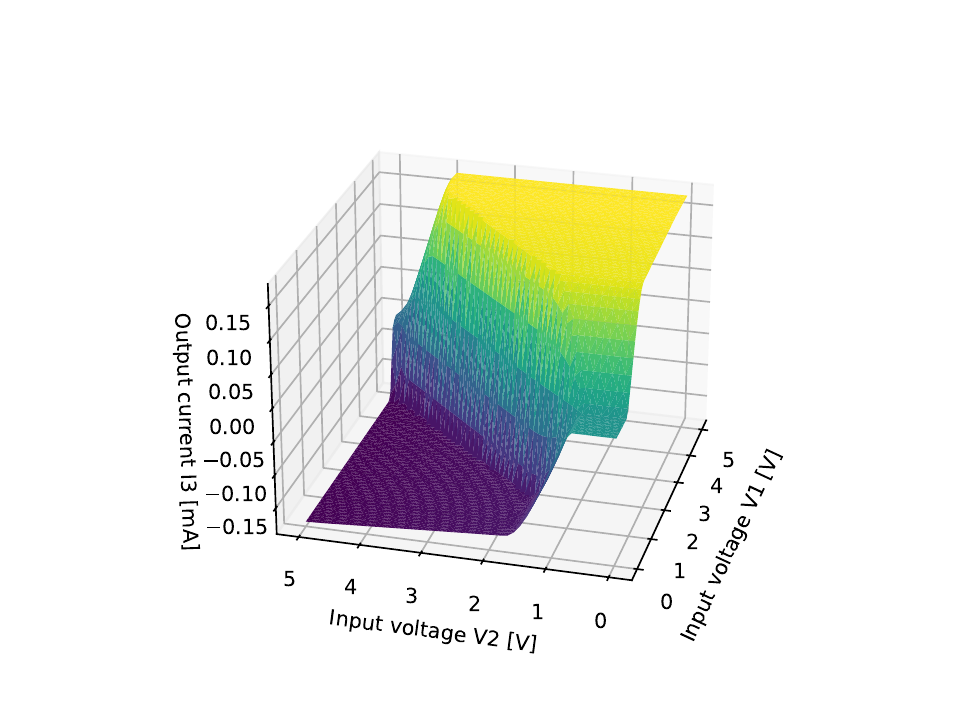}
      \caption{\small  DC output current $I^h_3(V_1,V_2)$ for $V_3$ = 3.5 V.}
      \label{fig:sub3}
    \end{subfigure}
    \caption{\small  Q3 interpolant surfaces for DC output current $I^h_3$ as a function of input voltages $(V_1,V_2)$ with fixed output voltage $V_3$.}
    \label{fig:DC_surface}
    
\end{figure}

\begin{figure}
    \centering
    \includegraphics[width=0.95\linewidth]{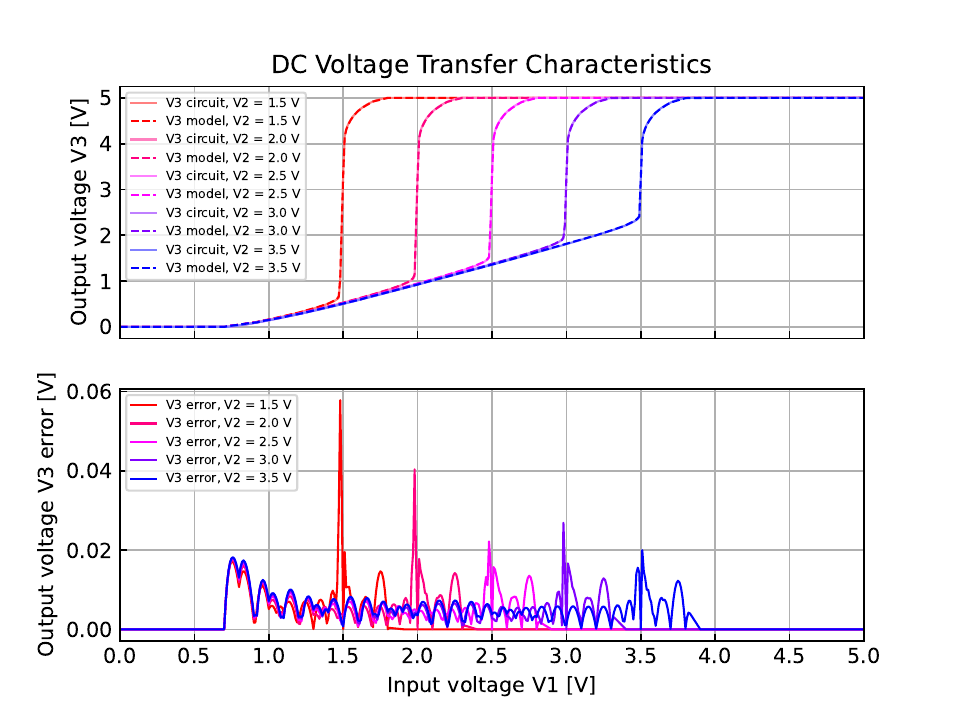}
    \caption{\small  \small Comparison between Xyce simulations and Hammerstein \mbox{\REV{DiffAmp}} model results of the DC voltage transfer characteristics of the differential amplifier. The output voltage $V^h_3$ is shown as a function of the input voltage $V_1$ for a fixed input voltage $V_2 \in \{1.5,2.0,3.5,3.0,3.5\}$ V.}
    \label{fig:DC_TF}
\end{figure}

The  plots in Figure \ref{fig:DC_surface} show the surfaces of $I^h_3(V_1,V_2)$ for \REV{two} different values of $V_3$. Figure \ref{fig:DC_TF} uses a two-dimensional format to compare the related output voltage $V^h_3(V_1,V_2)$ with the ``ground truth'' represented by the Xyce DC analysis simulation.  Xyce data are in the solid pastel colors and the data from the Hammerstein model are in the bold dashed colors. The error plots at the bottom of Figure \ref{fig:DC_TF} confirm that the Hammerstein model is in excellent agreement with the transistor-level Xyce simulation across different circuit operational points. The error spikes in the plots correspond to the regions where the term $|D^2 I_3|$ in estimate \eqref{eq:err-est-2} is large, i.e., the regions where $V_3$ and $I_3$ have large gradients. These spikes can be reduced by using non-uniform sampling that allocates more sampling points to these regions, while regions where $I_3$ and $V_3$ are ``flat'' are sampled more sparsely. Such an adaptive sampling strategy can also significantly reduce the memory cost of the trilinear interpolant. However, exploration of non-uniform sampling is beyond the scope of this paper.

%
%
%

\subsection{Transient results}\label{sec:tran_results}
In this subsection we use a number of  transient and alternating current (AC) simulations to compare predictions of the Hammerstein model with those of a transistor-level model of the circuit. We consider three instances of the model that all share the same input nonlinearity $\varphi$ but have dynamic blocks with increasing internal state dimensions $n=1,2,3$. Our goal is to demonstrate that the accuracy of the model increases with the state dimension, i.e., that $n$ can serve as a convenient ``knob'' to tune the quality of the transient simulations. To generate the ``ground truth'' data for these results we use Xyce to simulate the circuit with independent voltage sources connected to terminals 1 and 2 providing voltage waveforms $(v_1(t),v_2(t))$, and a 5 pF capacitive load connected to the output terminal. The voltage waveforms $(v_1(t),v_2(t))$ consist of various sinusoids at different frequencies, a square pulse, and the frequency-modulated training waveforms \eqref{eq:training_voltages} defined in Section \ref{sec:hammerstein}. The time-varying port currents $(i_1(t),i_2(t),i_3(t))$ are solved for within the simulation.
To generate the output of the Hammerstein model, we solve the closed-loop system of differential equations resulting from connecting the independent sources and capacitive load to the model
using the  Diffrax library \cite{Kidger_21_PHDTHESIS}. 
\paragraph{Reproductive tests}
\begin{figure}
    \centering
    \includegraphics[width=0.95\linewidth]{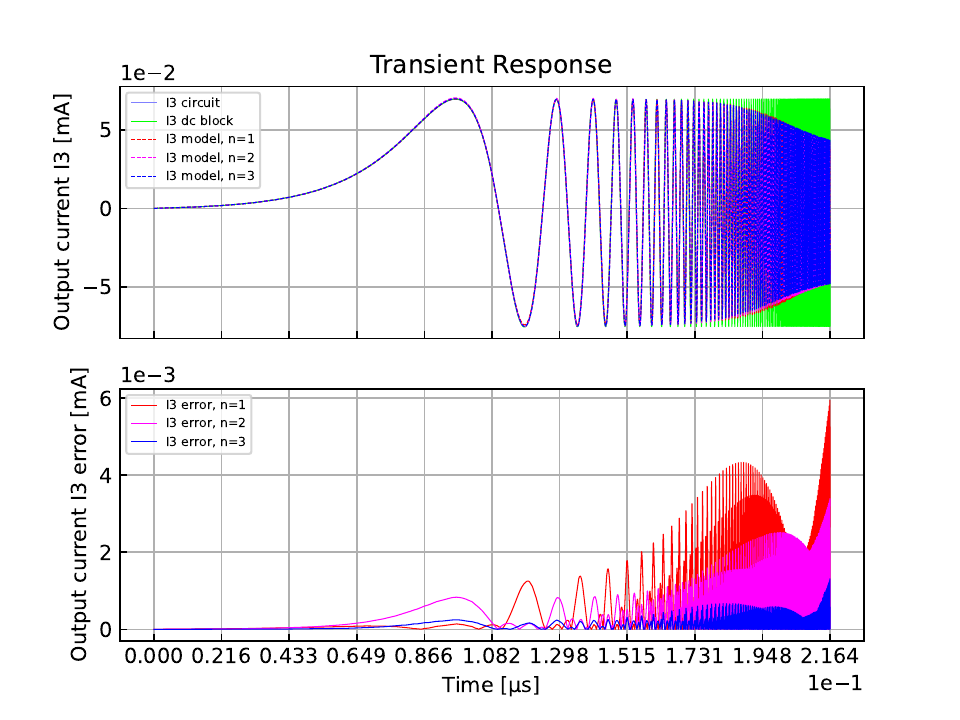}
    \caption{\small  Comparison of Hammerstein \REV{DiffAmp} models and Xyce simulation data for the training input voltage waveforms $v_1$ and $v_2$ given in equation \eqref{eq:training_voltages}.
    Top: transient output currents $i_3$. 
    Bottom: model output current error.}
    \label{fig:training}
    
\end{figure}
The plots in the Figure \ref{fig:training} compare transient analysis simulations of the Hammerstein \mbox{\REV{DiffAmp}} model and the transistor-level model in Xyce, performed for input voltages given by the training waveforms in equation \eqref{eq:training_voltages}. For comparison, the $I^h_3$ output of the nonlinear block $\varphi$ is also included to demonstrate the impact of eliminating the subsequent dynamic block. 
Since for low frequencies, the circuit is approximately memoryless, the accuracy of the Hammerstein \mbox{\REV{DiffAmp}} model is mainly based on the quality of the Q3 interpolant defining $\varphi$. As the frequency increases, we observe amplitude attenuation and phase shift, as well as subtle bias shift and distortion due to the interaction between the internal parasitic elements and the nonlinear amplification from the transistors. The top plot in Figure \ref{fig:training} clearly shows that these effects cannot be captured by the nonlinear block alone and that including the dynamic block improves the overall model fit, while the error plots at the bottom of the figure indicate that the accuracy of the model increases with the state dimension. In particular, we see that a model with a relatively small state dimension $n=3$ is already very effective in representing accurately the dynamical behaviors.

\begin{figure}
    \centering
    \includegraphics[width=0.95\linewidth]{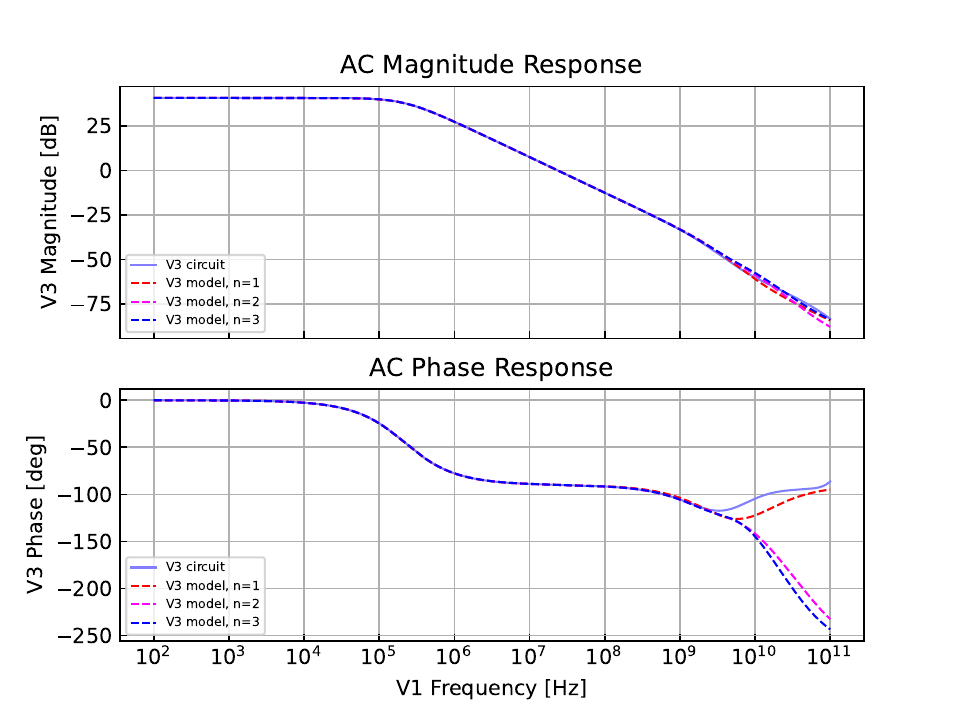}
    \caption{\small  Bode plot comparing the AC magnitude (top plot) and phase responses (bottom plot) between Xyce simulation of the circuit and the Hammerstein \mbox{\REV{DiffAmp}}  model. These results are computed with the amplifier set up in a single-input, single-output configuration $v_1 \mapsto v_3$, with $v_2 \equiv 0$.}
    \label{fig:ac}
    
\end{figure}

\begin{figure*}
    \centering
    \begin{subfigure}{0.5\textwidth}
        \centering
        \includegraphics[width=0.95\linewidth]{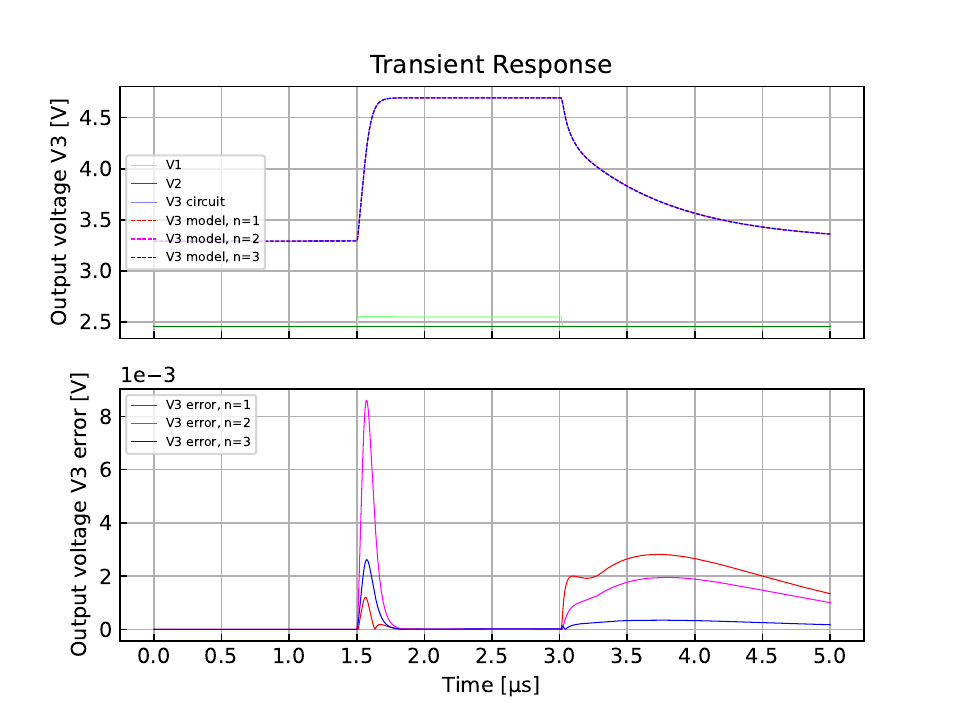}
        \caption{\small  Transient output voltage $v_3$ for a square wave input.}
    \end{subfigure}%
    \begin{subfigure}{0.5\textwidth}
        \centering
        \includegraphics[width=0.95\linewidth]{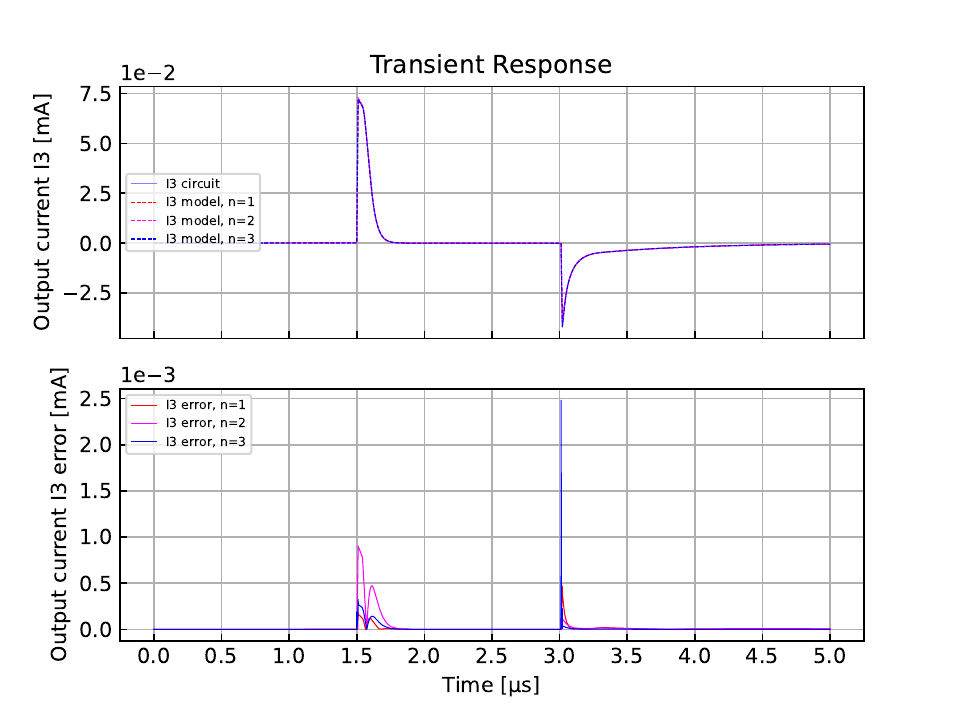}
        \caption{\small  Transient output current $i_3$ for a square wave input.}
    \end{subfigure}
    \caption{\small  Comparison of transient output voltages $v_3$ and currents $i_3$ between Xyce circuit simulations and Hammerstein \mbox{\REV{DiffAmp}} models for a square wave input voltage. The input voltage $v_1$ jumps between $2.45 V$ and $2.55 V$ with a ramp time of 10 ns (slew rate of $10 V / \SI{}{\micro\second}$); the input voltage $v_2$ is held at a constant $2.45 V$.}
    \label{fig:square}
\end{figure*}

\begin{figure*}
    \centering
    \begin{minipage}{0.5\textwidth}
    \begin{subfigure}{\textwidth}
        \centering
        \includegraphics[width=0.95\linewidth]{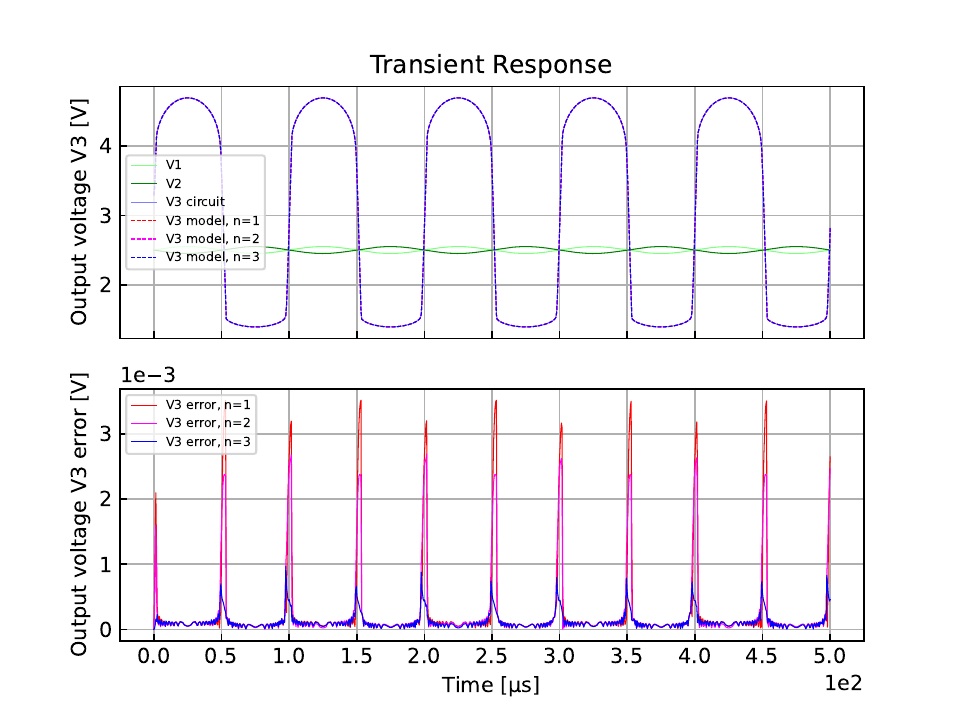}
        \caption{\small  Transient output voltage $v_3$ for input frequency $f = 10$ kHz.}
    \end{subfigure}
    \begin{subfigure}{\textwidth}
        \centering
        \includegraphics[width=0.95\linewidth]{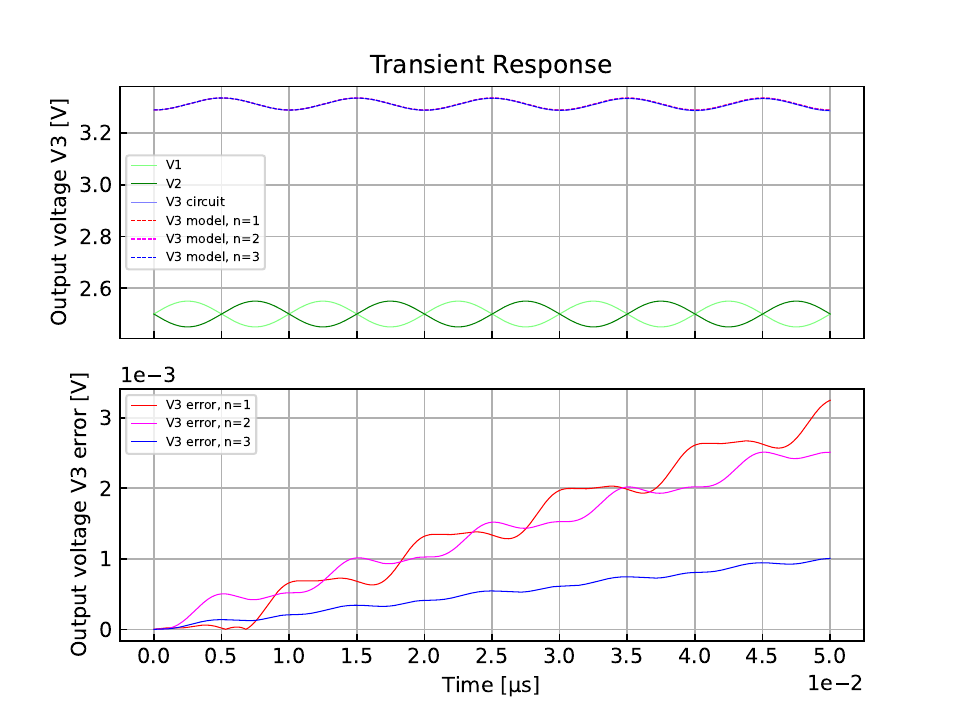}
        \caption{\small  Transient output voltage $v_3$ for input frequency $f = 100$ MHz.}
    \end{subfigure}
    \end{minipage}%
    \begin{minipage}{0.5\textwidth}
    \begin{subfigure}{\textwidth}
        \centering
        \includegraphics[width=0.95\linewidth]{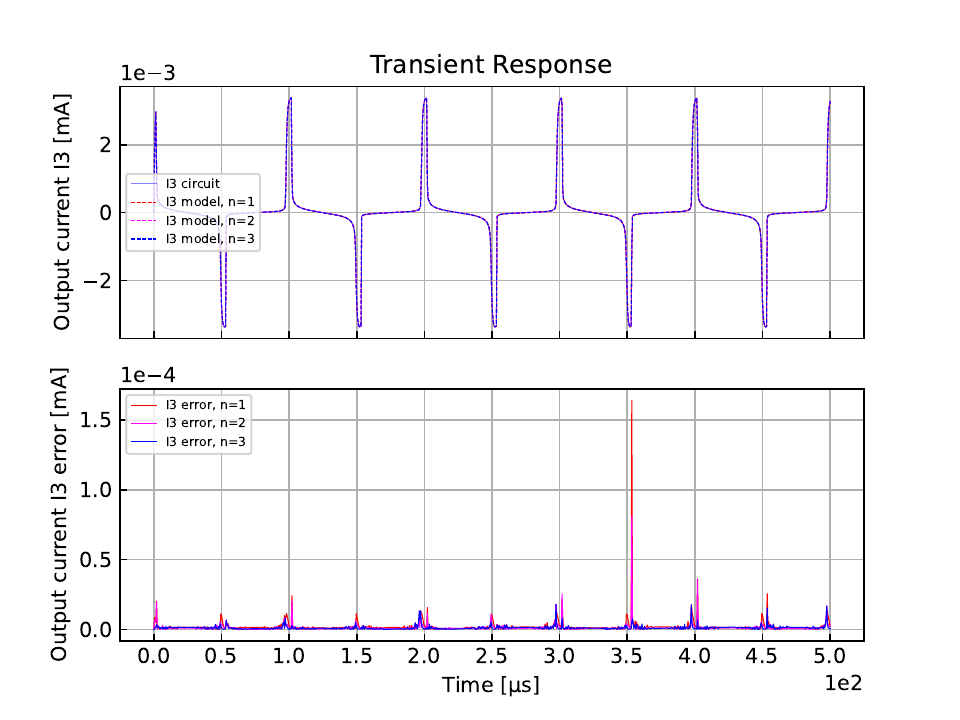}
        \caption{\small  Transient output current $i_3$ for input frequency $f = 10$ kHz.}
    \end{subfigure}
    \begin{subfigure}{\textwidth}
        \centering
        \includegraphics[width=0.95\linewidth]{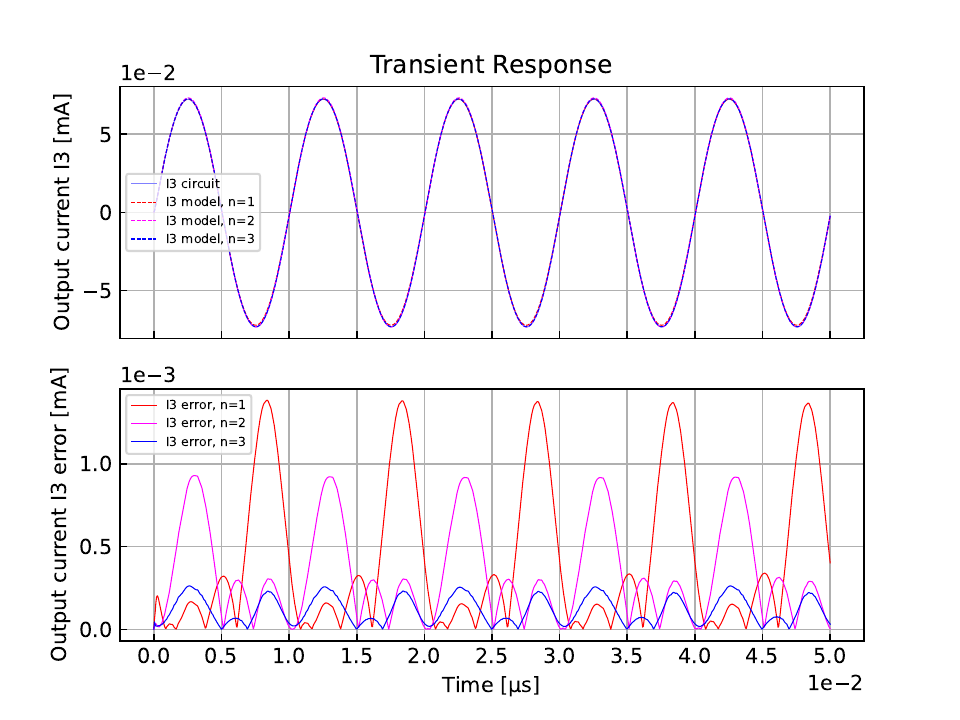}
        \caption{\small  Transient output current $i_3$ for input frequency $f = 100$ MHz.}
    \end{subfigure}
    \end{minipage}
    \caption{\small  Comparison of transient output voltages $v_3$ and currents $i_3$ between Xyce circuit simulations and Hammerstein \mbox{\REV{DiffAmp}} models for sinusoidal input voltages $v_1$ and $v_2$ with amplitude $A = 50$ mV, bias $V_{\text{bias}} = 2.5$ V, and frequencies (a,c) 10 kHz;  and (b,d) 100 MHz.}
    \label{fig:sinusoids}
\end{figure*}

\begin{figure*}[h!]
    \centering
    \begin{subfigure}{0.5\textwidth}
        \centering
        \includegraphics[width=0.95\linewidth]{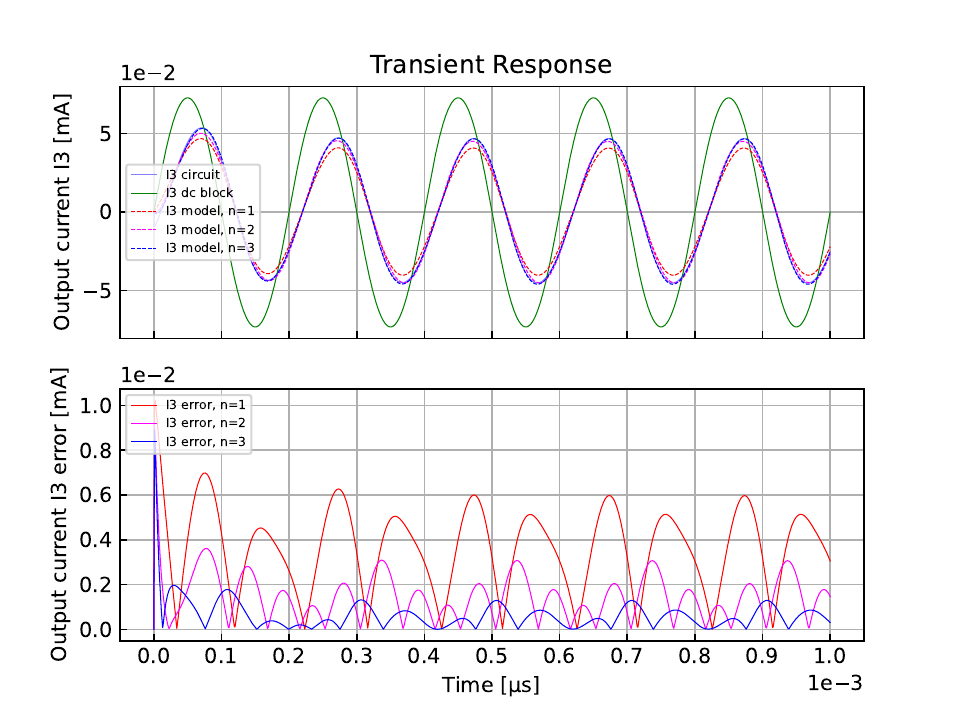}
        \caption{\small  Transient output current $i_3$ for input frequency $f = 5$ GHz.}
        \label{fig:sinusoids_high_freq-5}
    \end{subfigure}%
    \begin{subfigure}{0.5\textwidth}
        \centering
        \includegraphics[width=0.95\linewidth]{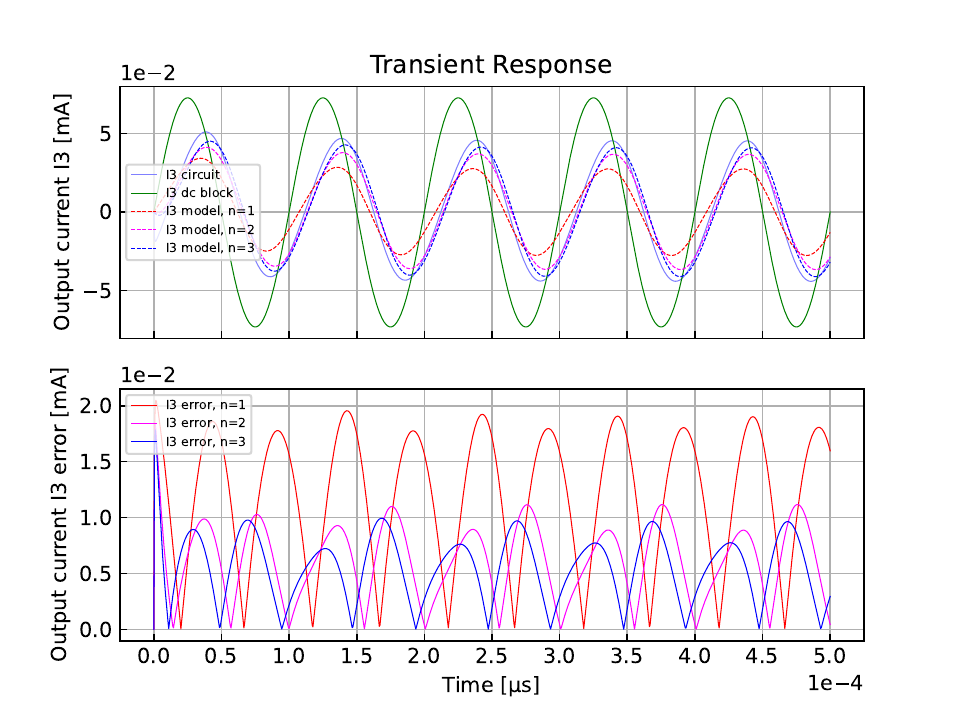}
        \caption{\small  Transient output current $i_3$ for input frequency $f = 10$ GHz.}
        \label{fig:sinusoids_high_freq-10}
    \end{subfigure}
    \caption{\small  Comparison of transient output voltages $v_3$ and currents $i_3$ between Xyce simulations and Hammerstein \mbox{\REV{DiffAmp}} models, including the DC current prediction $I_{\text{DC}}$, for high-frequency sinusoidal input voltages $v_1$ and $v_2$ with amplitude $A = 50$ mV and bias $V_{\text{bias}} = 2.5$ V. The model with state-dimension $n=3$ is accurate up to 5 GHz, but the accuracies of all models suffer at 10 GHz which is outside the range of frequencies represented in the training stimulus.}
    \label{fig:sinusoids_high_freq}
\end{figure*}

\paragraph{Predictive tests} 
These tests compare the Hammerstein \mbox{\REV{DiffAmp}} model with Xyce circuit simulations for inputs not included in the training data. Figure \ref{fig:ac} showcases the results from an AC voltage analysis with a capacitive load connected to the output terminal. This analysis performs a linearization around the nominal operating point $V_1 = V_2 = 2.5$ V and $V_3$ is computed as a root of the equation $0 = I_{\text{DC}}(V_1,V_2,V_3).$ The input voltage $v_2$ is held constant and the AC input voltage $v_1$ is a unit amplitude sine wave with variable frequency. The magnitude and phase of the resulting output voltage $v_3$ are shown. Notice that magnitude and phase responses between the circuit and the model begin to diverge for high frequencies after around 4 GHz; these data are absent in the training waveforms described in equation \eqref{eq:training_voltages}.

Figures \ref{fig:square} and \ref{fig:sinusoids} show the transient voltage and currents for square and sinusoidal input voltages, respectively. The two frequencies of the sinusoids are chosen to show the behavior before and after the cutoff frequency of the amplifier when connected to the 5 pF capacitive load.

The instantaneous frequency of the training waveform \eqref{eq:training_voltages} sweeps up to a maximum of 5 GHz. Figure \ref{fig:sinusoids_high_freq} examines what happens when the Hammerstein model is evaluated outside this range. We first evaluate the model for a sinusoidal input waveform with frequency $f = 5$ GHz, then with frequency $f = 10$ GHz. We then compare predictions of the ``full'' Hammerstein model to its ``truncated'' version comprising just the static nonlinear block and Xyce circuit simulations. The plots in Figure \ref{fig:sinusoids_high_freq-5} show that the accuracy of the ``full'' Hammerstein model remains reasonable for $f = 5$ GHz, while the ``truncated'' model misses the phase shift and the amplitude attenuation at this frequency. As before, the accuracy of the models is directly related to their state dimension and the model with $n=3$ is the most accurate. The plots in Figure \ref{fig:sinusoids_high_freq-10} show that model errors become even more pronounced at $f = 10$ GHz. For $n=1,2$ the models are too inaccurate for most circuit simulations, however for $n=3$ the model accuracy is reasonable, especially considering that this high-frequency data was not represented in the training stimulus.


\section{\REV{Simulation results with operational amplifier}}\label{sec:opamp-results}

\REV{This section presents simulation results for a Hammerstein model of a proprietary CMOS operational amplifier (OpAmp) containing 51 transistors. This amplifier is a rail to rail, Gm compensated, class AB output stage amplifier with cascode biasing and enable/disable circuitry. It is also considerably larger (in terms of number of internal devices) than the CMOS differential amplifier in Section \ref{sec:hammerstein}. 
%
The Hammerstein model for this circuit was constructed and identified following the same steps as for the  DiffAmp model in Section \ref{sec:hammerstein}. The analysis performed will also be similar to the analysis of the DiffAmp model in Section \ref{sec:results}, with differences noted.}

\REV{
To generate the synthetic data we again use the chirp waveform \eqref{eq:training_voltages} but with the following modified parameters: high frequency $f_1 = 100$ MHz,  low frequency $f_0 = 1$ kHz, and amplitude $A = 0.5$ mV. The bias $V_{\text{bias}} = 1.65$ V is the nominal DC operating point given by the midpoint of $V_{\text{DD}}$ and ground. The number of periods is chosen to be $N_{\text{per}} = 8685$. The time horizon is given by $T = N_{\text{per}} \frac{\ln(f_1) - \ln(f_0)}{f_1 - f_0} \approx 1$ ms.
We sample these waveforms using a variable time step in such a way that each period is sampled by 500 points.}

\subsection{Implementation and setup}\label{sec:opamp-impl}
\REV{We implemented the Hammerstein OpAmp model  in Python, with the transient simulations again utilizing the  Diffrax library \cite{Kidger_21_PHDTHESIS}. 
However, to identify  $A,B,C,D$ in \eqref{eq:hammerstein} here we use the {\tt ssest} function in MATLAB$^{TM}$ with the  Canonical Variate Algorithm \cite{Larimore1990CanonicalVA} option.} 
\REV{We again discard the gate currents $i_1$ and $i_2$ and instead only showcase the output voltage $v_3$ and output current $i_3$ resulting from (constant or time-varying) input voltages $v_1$ and $v_2$. We have the same terminal connectivities, but with a 50 pF capacitive load (in the transient results).}


\subsection{DC results}\label{sec:opamp_DC_results}
\begin{figure}
    \centering
      \includegraphics[trim={2cm 0.6cm 2cm 2cm},clip,width=0.95\linewidth]{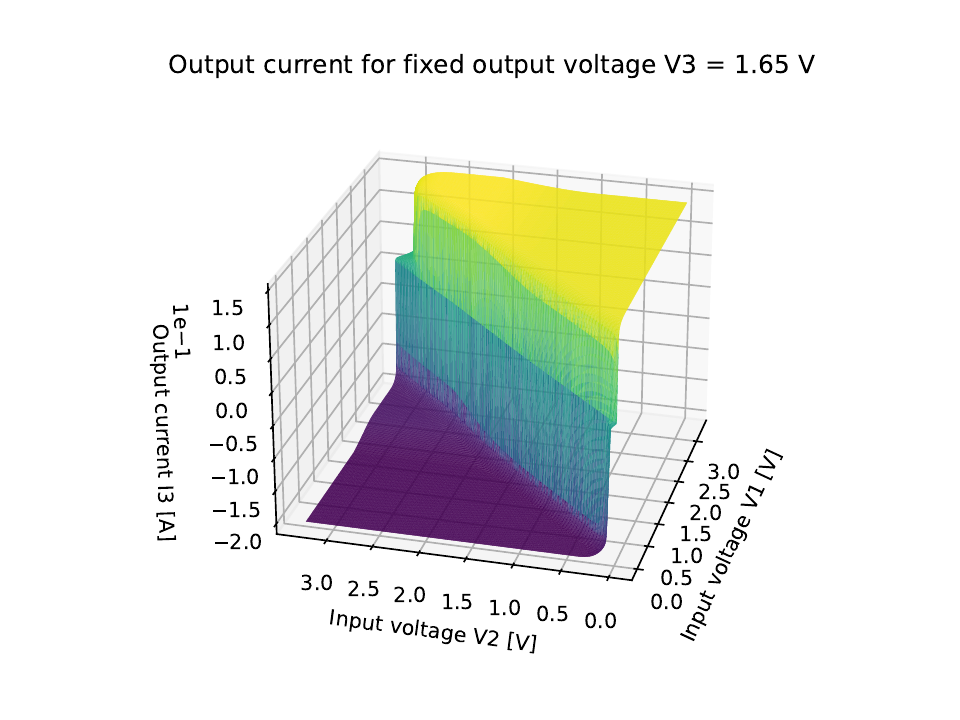}
    \caption{\small  Q3 interpolant surfaces for DC output current $I^h_3$ as a function of input voltages $(V_1,V_2)$ and a fixed output $V_3$ = 1.65V.}
    \label{fig:opamp_DC_surface}
    
\end{figure}
\begin{figure}
    \centering
        \includegraphics[width=0.95\linewidth]{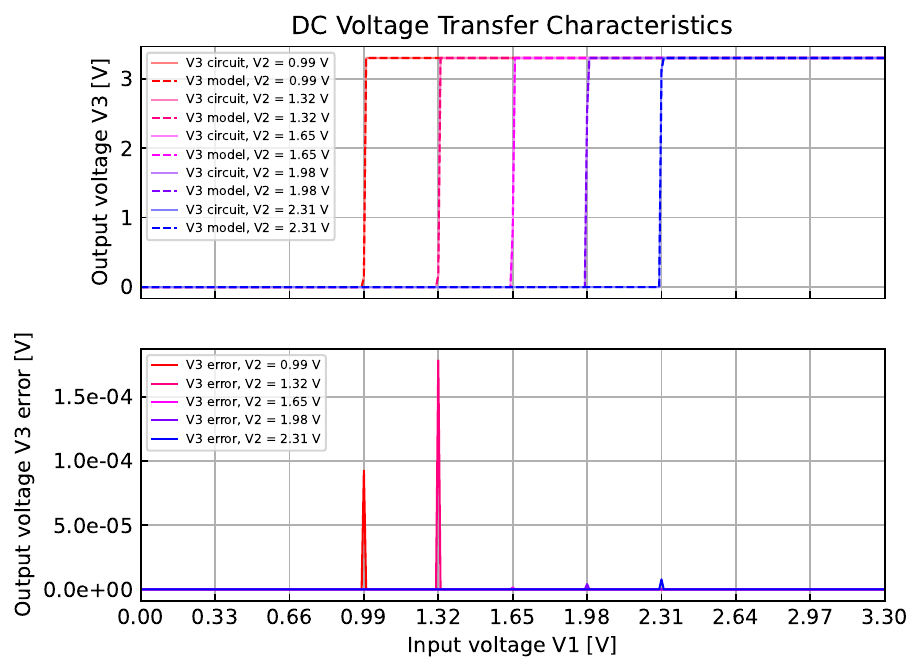}
    \caption{\small  Comparison between Xyce simulations and Hammerstein \REV{OpAmp} model results of the DC voltage transfer characteristics of the OpAmp. The output voltage $V^h_3$ is shown as a function of the input voltage $V_1$ for a fixed input voltage $V_2 \in \{0.99, 1.32, 1.65, 1.98, 2.31\}$ V.}
    \label{fig:opamp_DC_TF}
\end{figure}


\REV{Figure \ref{fig:opamp_DC_surface} show the surface of $I^h_3(V_1,V_2)$ for a fixed  value of $V_3$. Figure \ref{fig:opamp_DC_TF} uses a two-dimensional format to compare the related output voltage $V^h_3(V_1,V_2)$ with the Xyce ``ground truth''. The data from Xyce are in the solid pastel colors and the data from the Hammerstein \REV{OpAmp} model are in the bold dashed colors. The error plots at the bottom of Figure \ref{fig:opamp_DC_TF} confirm that the Hammerstein OpAmp model is in excellent agreement with the transistor-level Xyce simulations across different circuit operational points.} 
%


\subsection{Transient results}\label{sec:opamp-tran_results}
\REV{We use again a number of  transient and AC simulations to compare predictions of the Hammerstein \REVV{OpAmp} model with those of a transistor-level model of the circuit.} 

\paragraph{Reproductive tests}

\begin{figure}
    \centering
    \includegraphics[width=0.95\linewidth]{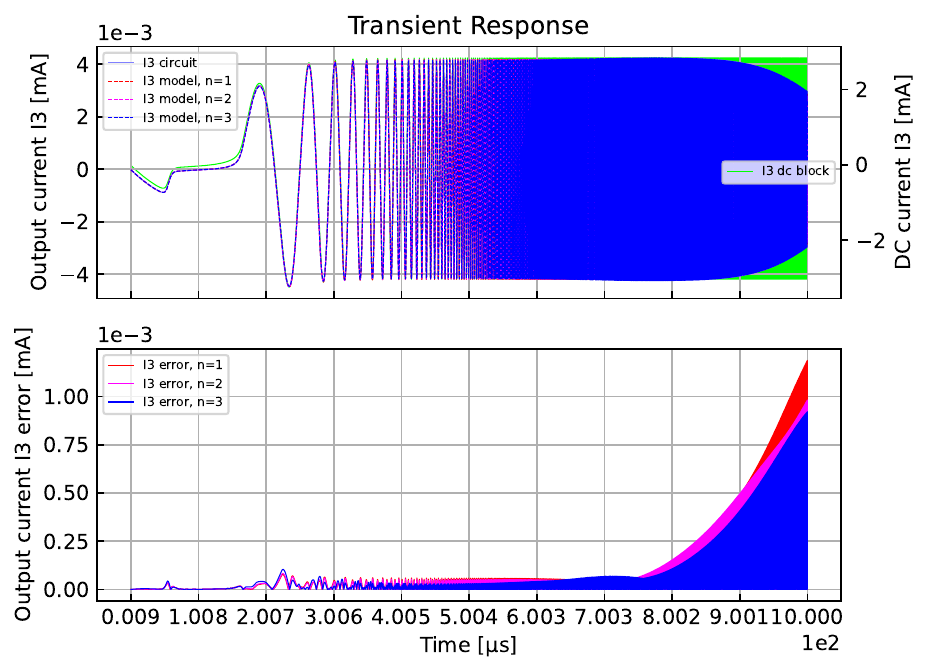}
    \caption{\small  Comparison of Hammerstein \REVV{OpAmp} models and Xyce simulation data for the training input voltage waveforms $v_1$ and $v_2$ 
    with frequency up to 100 MHz.
    Top: transient output currents $i_3$. 
    Bottom: model output current error.}
    \label{fig:opamp_training}
\end{figure}

\REV{The plots in the Figure \ref{fig:opamp_training} compare transient analysis simulations of the Hammerstein \REVV{OpAmp} model and the transistor-level model in Xyce. The top plot in Figure \ref{fig:opamp_training} shows that the nonlinear block alone cannot fully reproduce the amplitude attenuation and phase shift as the frequency of the input signal increases.}

\paragraph{Predictive tests} 

\begin{figure}
    \centering
    \includegraphics[width=0.95\linewidth]{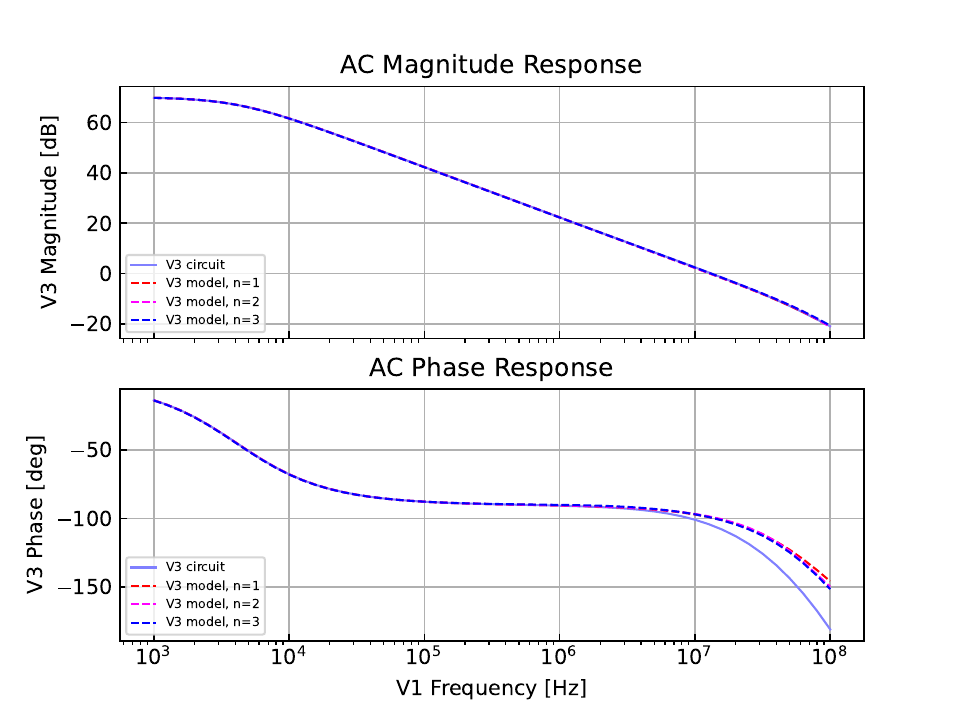}
    \caption{\small  Bode plot comparing the AC magnitude (top plot) and phase responses (bottom plot) between Xyce simulation of the circuit and the Hammerstein \REVV{OpAmp} model. These results are computed with the amplifier set up in a single-input, single-output configuration $v_1 \mapsto v_3$, with $v_2 \equiv 0$.}
    \label{fig:opamp_ac}
\end{figure}

\begin{figure*}
    \centering
    \begin{subfigure}{0.5\textwidth}
        \centering
        \includegraphics[width=0.95\linewidth]{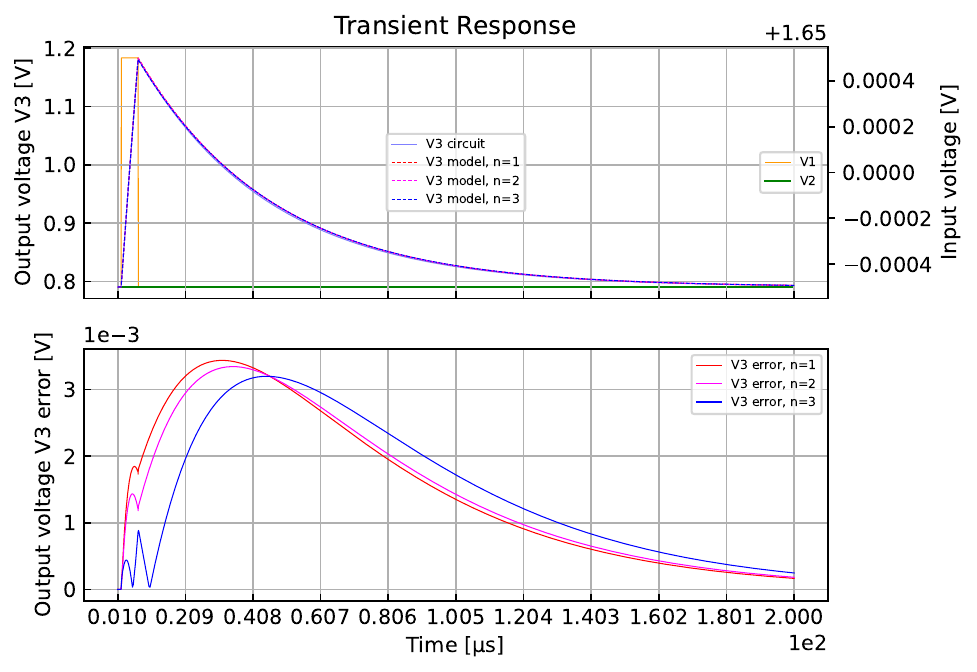}
        \caption{\small  Transient output voltage $v_3$ for a square wave input.}
    \end{subfigure}%
    \begin{subfigure}{0.5\textwidth}
        \centering
        \includegraphics[width=0.83\linewidth]{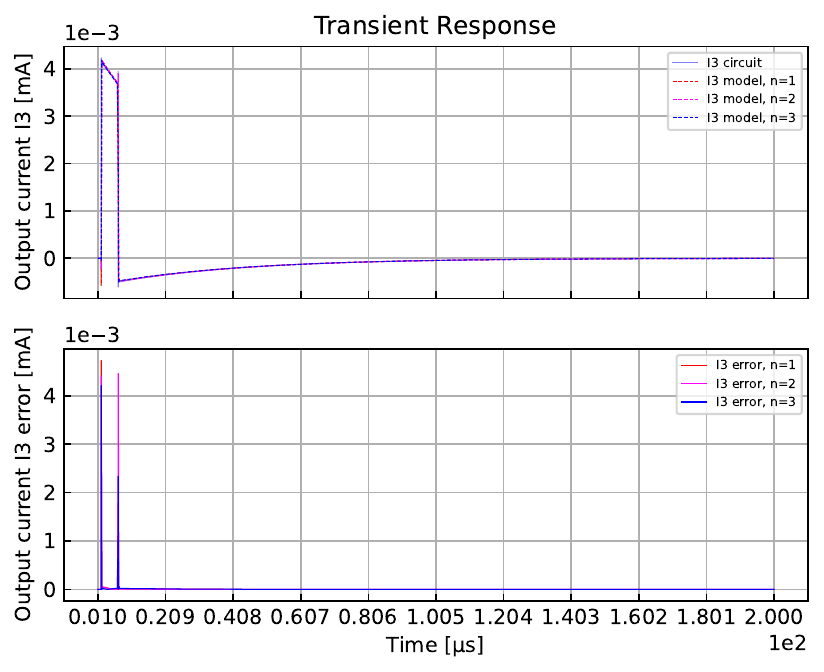}
        \caption{\small  Transient output current $i_3$ for a square wave input.}
    \end{subfigure}
    \caption{\small  Comparison of transient output voltages $v_3$ and currents $i_3$ between Xyce simulations and Hammerstein \mbox{\REV{OpAmp}} models for a square wave input voltage. The input voltage $v_1$ jumps between $1.6495 V$ and $1.6505 V$ with a ramp time of 10 ns (\REV{slew rate of $1 V / \SI{}{\micro\second}$}); the input voltage $v_2$ is held at a constant $1.65 V$.}
    \label{fig:opamp_square}
\end{figure*}

\begin{figure*}
    \centering
    \begin{minipage}{0.5\textwidth}
    \begin{subfigure}{\textwidth}
        \centering
        \includegraphics[width=0.95\linewidth]{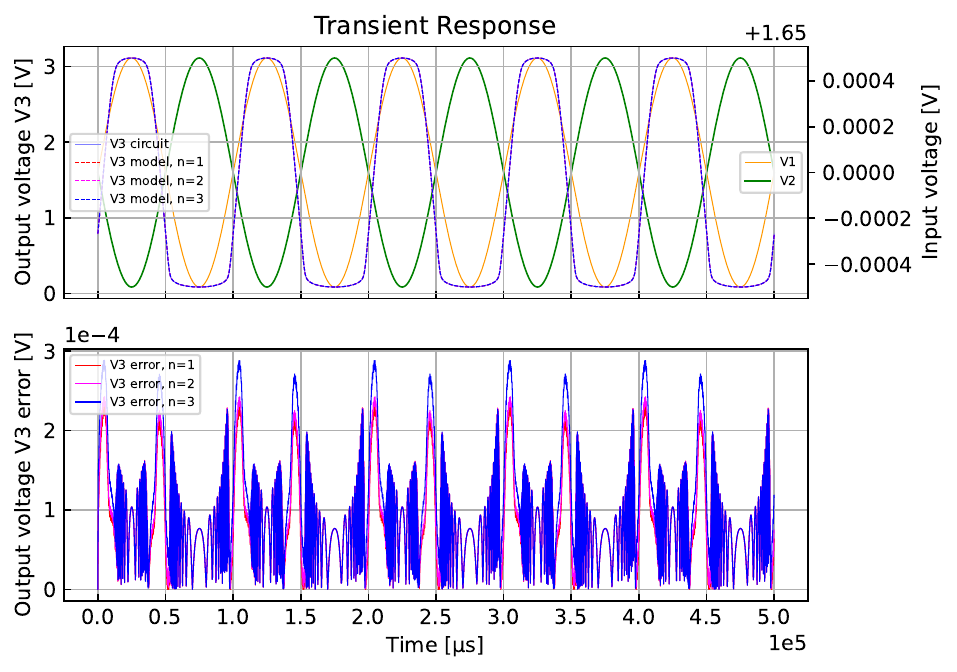}
        \caption{\small  Transient output voltage $v_3$ for input frequency $f = 10$ Hz.}
    \end{subfigure}
    \begin{subfigure}{\textwidth}
        \centering
        \includegraphics[width=0.95\linewidth]{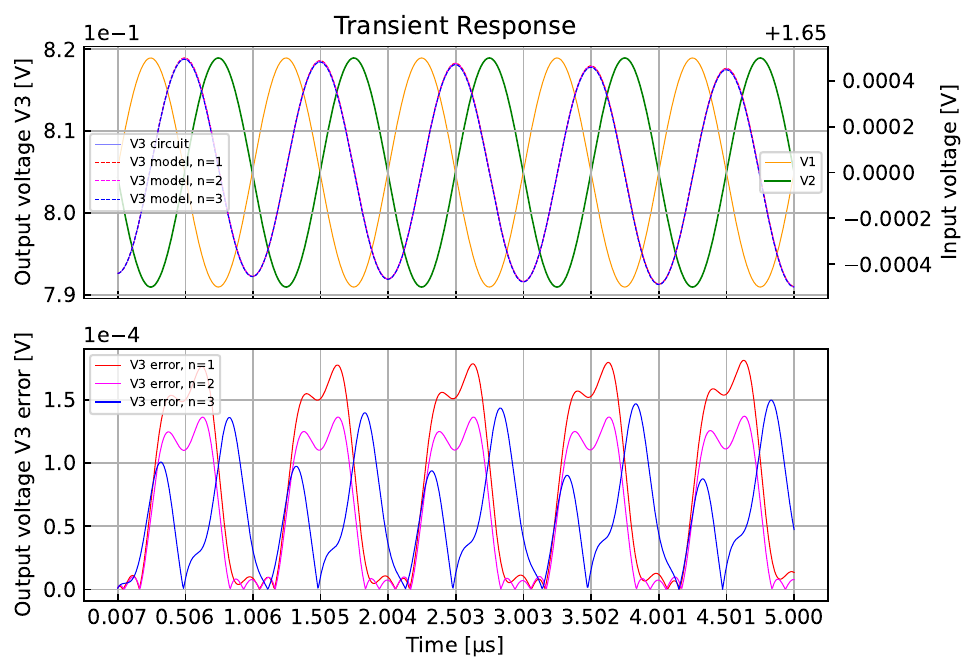}
        \caption{\small  Transient output voltage $v_3$ for input frequency $f = 1$ MHz.}
    \end{subfigure}
    \end{minipage}%
    \begin{minipage}{0.5\textwidth}
    \begin{subfigure}{\textwidth}
        \centering
        \includegraphics[width=0.85\linewidth]{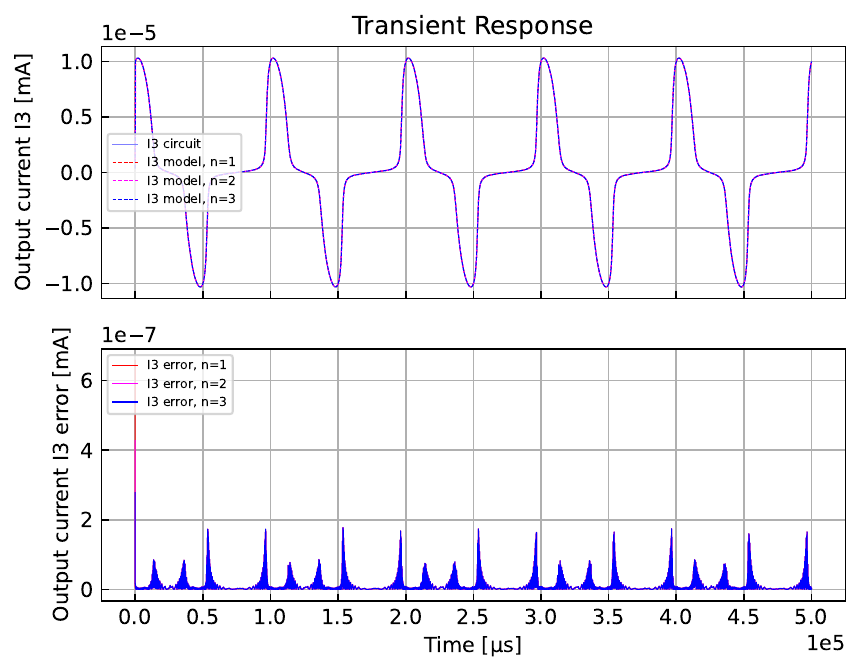}
        \caption{\small  Transient output current $i_3$ for input frequency $f = 10$ Hz.}
    \end{subfigure}
    \begin{subfigure}{\textwidth}
        \centering
        \includegraphics[width=0.85\linewidth]{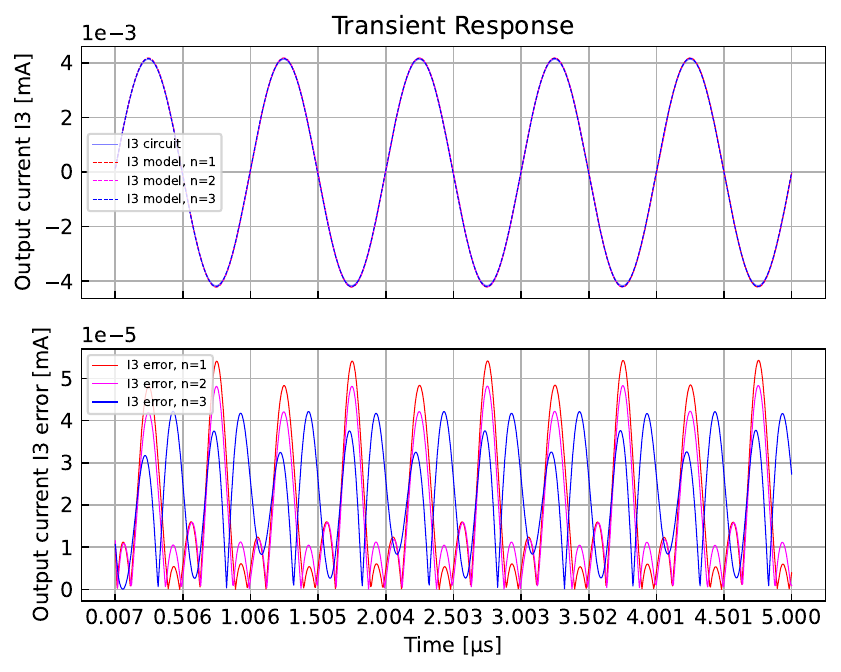}
        \caption{\small  Transient output current $i_3$ for input frequency $f = 1$ MHz.}
    \end{subfigure}
    \end{minipage}
    \caption{\small  Comparison of transient output voltages $v_3$ and currents $i_3$ between Xyce simulations and Hammerstein \mbox{\REV{OpAmp}} models for sinusoidal input voltages $v_1$ and $v_2$ with amplitude $A = 0.5$ mV, bias $V_{\text{bias}} = 1.65$ V, and frequencies (a,c) 10 Hz  and (b,d) 1 MHz.}
    \label{fig:opamp_sinusoids}
\end{figure*}

\begin{figure*}
    \centering
    \begin{subfigure}{0.5\textwidth}
        \centering
        \includegraphics[width=0.95\linewidth]{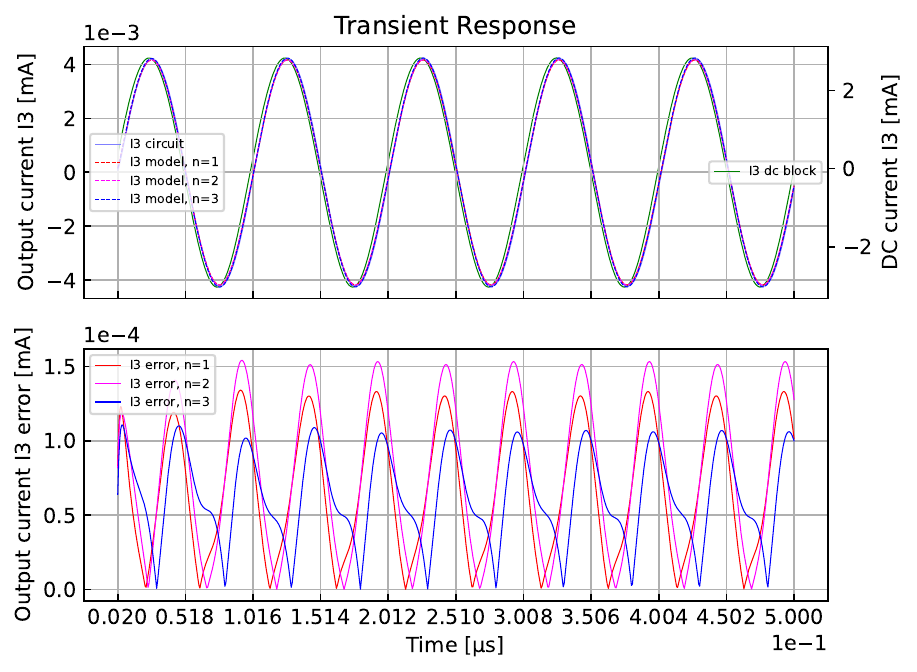}
        \caption{\small  Transient output current $i_3$ for input frequency $f = 10$ MHz.}
        \label{fig:opamp_sinusoids_high_freq-10}
    \end{subfigure}%
    \begin{subfigure}{0.5\textwidth}
        \centering
        \includegraphics[width=0.95\linewidth]{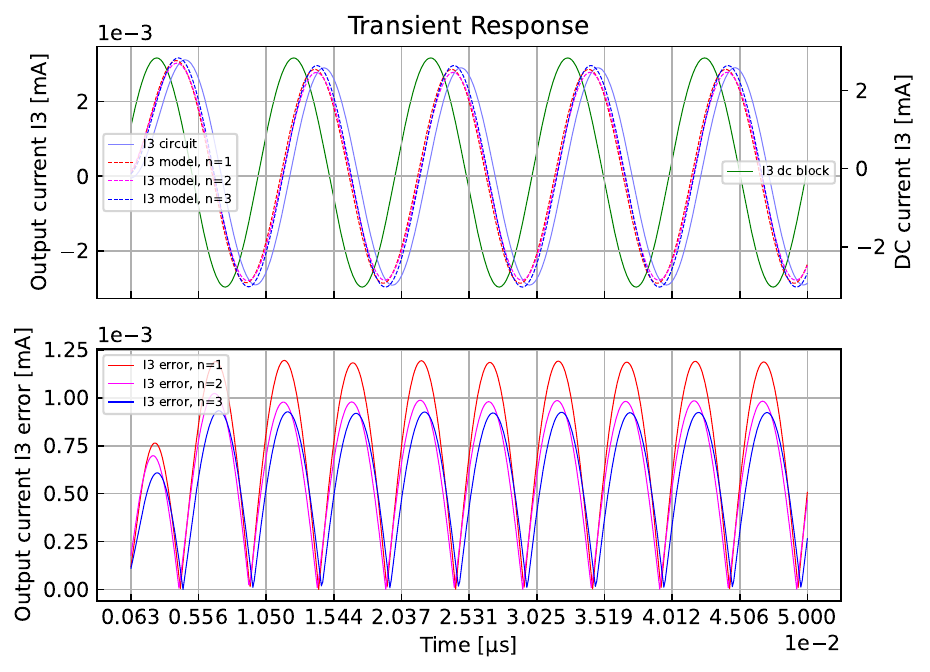}
        \caption{\small  Transient output current $i_3$ for input frequency $f = 100$ MHz.}
        \label{fig:opamp_sinusoids_high_freq-100}
    \end{subfigure}
    \caption{\small  Comparison of transient output voltages $v_3$ and currents $i_3$ between Xyce simulations and Hammerstein \mbox{\REV{OpAmp}} models, including the DC current prediction $I_{\text{DC}}$, for high-frequency sinusoidal input voltages $v_1$ and $v_2$ with amplitude $A = 0.5$ mV and bias $V_{\text{bias}} = 1.65$ V. }
    \label{fig:opamp_sinusoids_high_freq}
\end{figure*}

\REV{We compare the Hammerstein model with Xyce circuit simulations for inputs that have not been seen during the training process. Figure \ref{fig:opamp_ac} showcases the results from an AC voltage analysis with a capacitive load connected to the output terminal. This analysis performs a linearization around the nominal operating point $V_1 = V_2 = 1.65$ V and $V_3$ is computed as a root of the equation $0 = I_{\text{DC}}(V_1,V_2,V_3).$ The input voltage $v_2$ is held constant and the AC input voltage $v_1$ is a unit amplitude sine wave with variable frequency. The magnitude and phase of the resulting output voltage $v_3$ are shown. }
\REV{Figures \ref{fig:opamp_square} and \ref{fig:opamp_sinusoids} show the transient voltage and currents for square and sinusoidal input voltages, respectively. The two frequencies of the sinusoids are chosen to show the behavior before and after the cutoff frequency of the amplifier when connected to the 50 pF capacitive load.}

\REV{The instantaneous frequency of the training waveform sweeps from 1 kHz to 100 MHz. Figure \ref{fig:opamp_sinusoids_high_freq} examines what happens when the Hammerstein OpAmp model is evaluated near the boundary of the frequency range represented in the training stimulus. We first evaluate the model for a sinusoidal input waveform with frequency $f = 10$ MHz, then with frequency $f = 100$ MHz. We then compare predictions of the ``full'' and  ``truncated'' models with Xyce circuit simulations. The plots in Figure \ref{fig:opamp_sinusoids_high_freq-10} show that the accuracy of the ``full''  model remains reasonable for $f = 10$ MHz, while the ``truncated'' DC model misses the phase shift and the amplitude attenuation at this frequency. As before, the accuracy of the models is directly related to their state dimension and the model with $n=3$ is the most accurate.}

\REV{The plots in Figure \ref{fig:opamp_sinusoids_high_freq-100} show how the model accuracy degrades at $f = 100$ MHz. \REV{For all values of $n$, we observe that the models capture the output current amplitude but suffer a phase discrepancy at this frequency, which is near the top of the frequency range represented in the training stimulus.}}

\section{Discussion and conclusions}\label{sec:discussion}

\REV{The key contributions of this work are two-fold:}

\textbf{(1)} \REV{We demonstrated that parametric SysID can be used as an effective MOR approach, and the sequential model inference for Hammerstein architectures formulated in this paper is an effective training strategy that enables a simplified model development workflow.}

\textbf{(2)} \REV{Our simulation results confirm and substantiate that the set of system behaviors expressible by a Hammerstein architecture is rich enough to accurately and efficiently represent some relevant and versatile classes of nonlinear circuits, as demonstrated by the Hammerstein models of the differential and operational amplifiers formulated and studied in this work.}

\REV{The contribution (2) is of important practical interest because the Hammerstein model is among the simplest possible nonlinear model forms in the system hierarchy ranging from linear systems (whose admissible behaviors are essentially completely characterized) to universal nonlinear architectures (such as non-parametric data-driven models, and dynamical models based on universal function classes like neural nets or polynomial expansions).
It also reconfirms that choosing simpler (yet still sufficiently expressive) architectures in general yields more computationally efficient models that are also easier to train, requiring less training data and/or fewer optimization epochs to achieve convergence.}

\REV{Notably, our results indicate that simplicity of the model form does not inhibit its ability to capture the salient circuit behaviors using a small number of internal states. For example, reproductive and predictive tests of Hammerstein models with state dimensions $n\le 3$ reveal excellent fit with transistor-level simulations performed in Xyce.} \REVV{We note that these models comprise systems of only $n\le 3$ ODEs, whereas the original DiffAmp and OpAmp circuits require solving a system of DAEs with 24 and 172 independent solution variables, respectively.} We expect that comparable or better model compression factors can be achieved for other complex circuits with similar ``scripted'' behaviors (e.g., op amps and comparators) but large device counts, \REVV{providing a foundation for computationally efficient system-level analysis and rapid design assessments, and evaluations.} Such circuits often have highly nonlinear mathematical descriptions that are difficult to handle by conventional, structure-dependent, \REV{and/or} intrusive MOR approaches, thus are good candidates for data-driven behavioral modeling approaches.

\section*{Acknowledgments}

\footnotesize {The authors would like to thank Sean Pearson from Sandia National Laboratories for his guidance and advice with the analysis of the operational amplifier circuit. This work was supported by the Sandia National Laboratories (SNL) Laboratory-directed Research and Development (LDRD) program, and the U.S. Department of Energy, Office of Science, and Office of Advanced Scientific Computing Research under Award Number  DE-SC-0000230927. Sandia National Laboratories is a multimission laboratory managed and operated by National Technology and Engineering Solutions of Sandia, LLC., a wholly owned subsidiary of Honeywell International, Inc., for the U.S. Department of Energy's National Nuclear Security Administration contract number DE-NA0003525. \REVV{This written work is authored by an employee of NTESS. The employee, not NTESS, owns the right, title and interest in and to the written work and is responsible for its contents. Any subjective views or opinions that might be expressed in the written work do not necessarily represent the views of the U.S. Government. The publisher acknowledges that the U.S. Government retains a non-exclusive, paid-up, irrevocable, world-wide license to publish or reproduce the published form of this written work or allow others to do so, for U.S. Government purposes. The DOE will provide public access to results of federally sponsored research in accordance with the DOE Public Access Plan.}}


\bibliographystyle{plain}
\bibliography{R2R}

\section*{Biographies}

\begin{IEEEbiography}[{\includegraphics[width=1in,height=1.25in,clip,keepaspectratio]{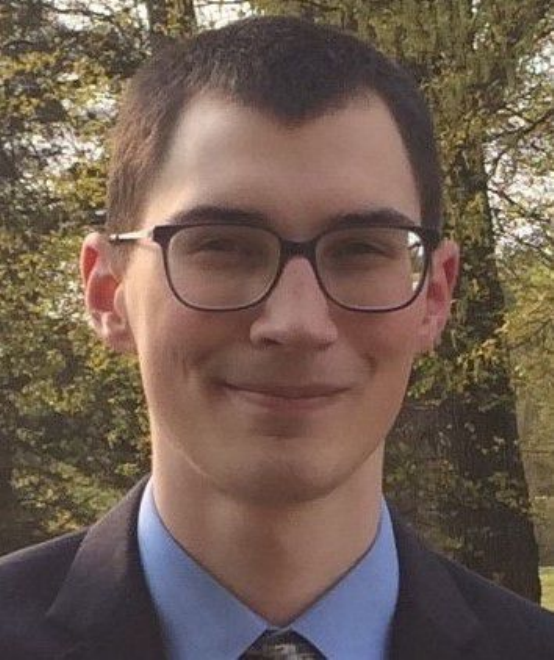}}]{Joshua Hanson}
Joshua is a Ph.D. candidate in Electrical Engineering at the University of Illinois at Urbana-Champaign. He earned his B.S. in Electrical Engineering and B.S. in Physics in 2018 at Clemson University, his M.S. in Electrical Engineering in 2020 and M.S. in Mathematics in 2021 at the University of Illinois, and joined Sandia National Laboratories as a student intern in 2020. His research interests include modeling and simulation, system identification, and control, with a focus on data-driven methods and scientific machine learning.
\end{IEEEbiography}
\vspace{-1cm}
\begin{IEEEbiography}[{\includegraphics[width=1in,height=1.25in,clip,keepaspectratio]{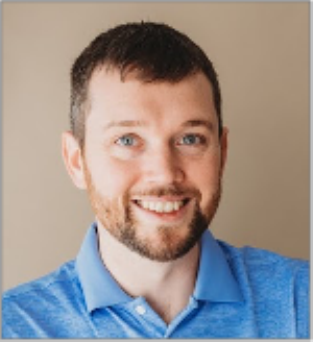}}]{Paul Kuberry}
Paul is a Principal Member of Technical Staff in the Computational Mathematics Department at Sandia National Laboratories. He received his B.S. in Mathematics from Clarion University of Pennsylvania in 2010 and his M.S. and Ph.D. in Mathematical Sciences from Clemson University in 2012 and 2015. Paul is a software developer for Xyce-PyMi, a Python model interpreter for the circuit simulator Xyce. He is a computational scientist developing novel techniques for data-driven device models and their integration in Spice-like simulators.
\end{IEEEbiography}
\vspace{-1cm}
\begin{IEEEbiography}[{\includegraphics[width=1in,height=1.25in,clip,keepaspectratio]{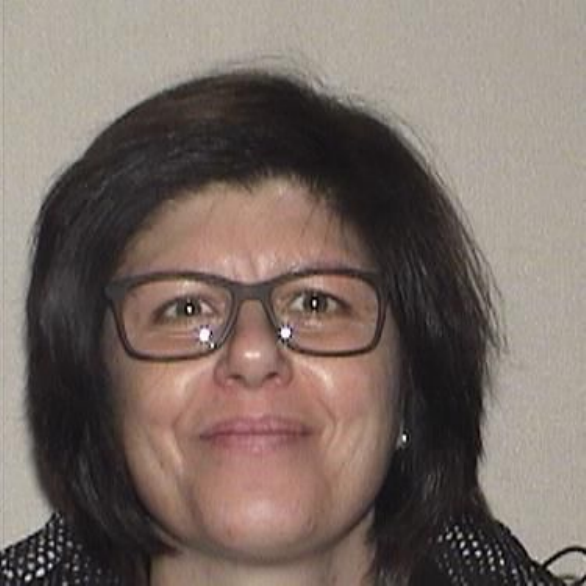}}]{Biliana Paskaleva}
Biliana is a Principal Member of Technical Staff in the Radiation Modeling and Analysis Department at Sandia National Laboratories. She earned her Ph.D. in Electrical and Computer Engineering from the University of New Mexico in 2009 and joined Sandia in 2010. Biliana's interests include validation and verification of electrical systems and devices, multispectral and hyperspectral image classification, and data-driven compact models for semiconductor devices.
\end{IEEEbiography}
\vspace{-1cm}
\begin{IEEEbiography}[{\includegraphics[width=1in,height=1.25in,clip,keepaspectratio]{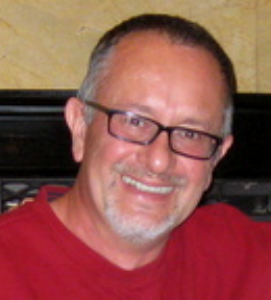}}]{Pavel Bochev}
Pavel is a Senior Scientist at the Center for Computing Research at Sandia National Laboratories. He earned his Ph.D. in Mathematics in 1994 at Virginia Tech and has been with Sandia National Laboratories since 2000. His research interests include numerical analysis and applied mathematics, with particular emphasis on compatible discretization methods for PDEs, meshfree methods, optimization and control problems, and scientific machine learning.
\end{IEEEbiography}

\end{document}